\documentclass[prb,twocolumn,showpacs,superscriptaddress,preprintnumbers,floatfix]{revtex4}

\usepackage{dcolumn}
\usepackage{amsmath}
\usepackage{amssymb}
\usepackage{graphicx}
\usepackage{bm}   

\newcommand{\eM}     {$\epsilon$-machine}
\newcommand{\eMs}    {$\epsilon$-machines}
\newcommand{\EM}     {$\epsilon$-Machine}
\newcommand{\EMs}    {$\epsilon$-Machines}

\newcommand{\CausalState}	{ {\cal S} }

\newcommand{\causalstate}	{ \sigma }
\newcommand{\CausalStateSet}	{ \boldsymbol{\CausalState} }

\newcommand{\Prob}		{ {\rm P}}

\newcommand{\Cmu}		{ {C_\mu}}
\newcommand{\hmu}		{ {h_\mu}}
\newcommand{\EE}		{ {\bf E}}

\newcommand{\ProcessAlphabet}	{\mathcal{A}}

\newcommand{\Range} {{r}}
\newcommand{\CSCP} {{P_{CSC}}}
\newcommand{\eMSR}    {{{$\epsilon$}MSR}}
\newcommand{\MLength}  {\Range_l}
\newcommand{\InteractionLength}  {\Range_I}

\begin{document}


\title{Inferring Pattern and Disorder in\\
Close-Packed Structures from X-ray Diffraction Studies, Part I:\\
\EM\ Spectral Reconstruction Theory}

\author{D. P. Varn}
\affiliation{Santa Fe Institute, 1399 Hyde Park Road, Santa Fe, New Mexico 87501}
\affiliation{Department of Physics and Astronomy,
University of Tennessee, Knoxville, Tennessee 37996}

\author{G. S. Canright}
\affiliation{Department of Physics and Astronomy, University of Tennessee,
Knoxville, Tennessee 37996}
\affiliation{Telenor Research and Development, 1331 Fornebu, Norway}

\author{J. P. Crutchfield}
\affiliation{Santa Fe Institute, 1399 Hyde Park Road, Santa Fe, New Mexico 87501}

\date{\today}

\begin{abstract}
In a recent publication [D. P. Varn, G. S. Canright, and J. P. Crutchfield,
Phys. Rev. B {\bf 66}:17, 156 (2002)] we introduced a new technique for discovering
and describing planar disorder in close-packed structures (CPSs) directly
from their diffraction spectra. Here we provide the theoretical development
behind those results, adapting computational mechanics to describe
one-dimensional structure in materials. By way of contrast, we give a detailed
analysis of the current alternative approach, the fault model (FM), and offer
several criticisms. We then demonstrate that the computational mechanics
description of the stacking sequence---in the form of an
\eM---provides the minimal and unique description of the crystal, whether
ordered, disordered, or some combination. We find that we can detect and
describe any amount of disorder, as well as materials that are mixtures of
various kinds of crystalline structure. For purposes of comparison, we show
that in some special limits it is possible to reduce the \eM\ to the FM's
description of faulting structures. The comparison demonstrates that an \eM\
gives more physical insight into material structures and also more accurate
predictions of those structures. From the \eM\ it is possible to calculate
measures of memory, structural complexity, and configurational entropy.
We demonstrate our technique on four prototype systems and find that it
provides stacking descriptions that are superior to any so far used in the
literature. Underlying this approach is a novel method for \eM\ reconstruction
that uses correlation functions estimated from diffraction spectra, rather
than sequences of microscopic configurations, as is typically used in other
domains. The result is that the methods developed here can be adapted to a
wide range of experimental systems in which spectroscopic data is available.

\pacs{
  61.72.Dd,   
  61.10.Nz,   
  61.43.-j,   
  81.30.Hd    
  \\
  \begin{center}
  Santa Fe Institute Working Paper 03-02-XXX
  ~~~~~~~~arxiv.org/cond-mat/0302528
  \end{center}
  }

\end{abstract}

\maketitle

\section{\label{sec:level1}Introduction}

Fundamental to understanding the physical properties of a solid is a
thorough description of its composition and the arrangement of its
constituent parts. While chemical analysis provides information about
composition, many complementary methods---such as x-ray diffraction, electron
diffraction, high-resolution electron microscopy, optical microscopy, and
x-ray diffraction tomography---are essential to discovering the large-scale
structure of crystalline materials. For example, the placement of Bragg peaks
in x-ray diffraction spectra has proved to be a particularly powerful
source of structural information in ordered solids for nearly a
century.~\cite{l18} For solids that deviate from a strict periodic
ordering of their constituent atoms, however, the diffraction spectrum
typically shows a weakening and broadening the Bragg peaks as well as
the appearance of diffuse scattering. As the disorder becomes more
pronounced, the Bragg peaks disappear altogether and leave a completely
diffuse spectrum. The problem of inferring crystal structure for such
disordered materials from x-ray diffraction has been addressed by many
researchers.\cite{f95,j87,pw98,w85,s82,g63} In the most general case,
however, the problem remains unsolved. Indeed, it is known that without
the assumption of strict crystallinity, the problem has no unique
solution.~\cite{w97}

Many kinds of disorder are present in solids,~\cite{hl82,am76,k96}
such as Schottky defects, substitution impurities, screw and edge
dislocations, and planar slips. Of these, only planar slips will be
considered here. Planar defects occur in crystal structure when one
crystal plane is displaced from another by a non-Bravais lattice vector.
These slips can occur during crystal growth or can result from some
stress to the crystal, be it mechanical, thermal, or even through
irradiation.~\cite{sk94} When an otherwise perfect crystal has some
planar disorder, the portion of the structure that cannot be thought
of as part of the crystal is called a \emph{fault}.  If there is a
transition between two crystal structures, the interface between the
two is also known as a fault, even if each layer can be thought of
as belonging to one of the crystal structures.  Many different kinds of
faults have been postulated, including growth faults, deformation faults,
and layer-displacement faults.~\cite{sk94,f51a,fb81}

Planar defects are surprisingly common in crystals, being especially
prevalent in a broad class of materials know as
\emph{polytypes}.\cite{vk66,sk94,t91,pk82,fps92} First discovered in SiC by
Baumhauer~\cite{b12} in 1912, polytypism has since been found in dozens
of materials. Polytypism is the phenomenon of solids built up from
identical~\cite{note1} layers, called \emph{modular layers}
(MLs),~\cite{vc01} that differ only in the manner of the
stacking. Typically, one finds that the \emph{intra}-ML interactions
are relatively strong as compared to the \emph{inter}-ML interactions,
so that disorder \emph{within} a ML is rare.  Energetic considerations
usually restrict the allowed orientations of MLs to a discrete set, with
only a small energy difference between two different stackings.  Thus,
the description of a polytype, ordered or disordered, formally reduces
to a one-dimensional list---called the \emph{stacking sequence}---that
lists successive orientations encountered as one moves along the stacking
direction.

The small energy difference between different stackings arises because
the coordination of the nearest neighbor, next-nearest neighbor, and
sometimes even higher neighbors is often the same regardless of the
stacking arrangement. It is therefore possible to have many distinct
stackings---some periodic and some not. For several of the most polytypic
materials---e.g., SiC, ZnS, and CdI$_2$---there are about $150$, $185$, and
$200$ known crystalline structures, respectively. Remarkably, some have unit
cells extending over 100 MLs.~\cite{sk94} The stacking period of many such
polytypes is far in excess of the calculated inter-ML interactions,
which are estimated to be, for example, $\sim 1$ ML in ZnS~\cite{en90} and
$\sim 3$ ML in SiC.~\cite{cnhc87,sh90} Nearly a dozen theories have been
proposed,~\cite{sk94,t91,f51b,pk82,j54a,y88,p89} yet a satisfactory and
systematic explanation is still lacking for the diversity and kinds of
observed structure. 

Much of the interest in polytypism has centered around the issue
of long-range order and the existence of so 
many apparently stable structures. Reconciling the calculated 
range of interaction between MLs with the length scale over which 
organization appears has been the chief mystery of polytypism. Also of 
interest is the characterization of the solid-state transformations
common in many of these materials. While the length scale on which spatial
organization appears in these materials is easily found for crystal structures,
the similar question for disordered structures has not so far been addressed.
What is needed is a model is that gives a statistical description of the
observed stacking sequences from which characteristic length parameters  
are calculable. Finally, from a unified description of both crystalline and 
noncrystalline structures, a more comprehensive picture of polytypism should
emerge and hopefully render polytypism more amenable to theoretical discussion
and analysis. 

A significant source of information about the structure of solids is derived
from diffraction spectra. While it can be challenging to identify crystal
structures with units cells over 100s of MLs, in fact most periodic
structures have been identified.~\cite{sk94} In many polytypes, disordered
sequences are also common, and a single crystal can contain regions of both
ordered and disordered stackings. A main goal of this present work is to
develop a technique for discovering and describing planar disorder in
close-packed structures (CPSs)
from diffraction spectra. A further task is to detail the connection between
our model of disordered structures and physically relevant parameters derivable
from it. We will then be in a position to treat similar questions about the
possibility of long-range order in disordered structures.

\subsection{Indirect Methods}

The problem of quantifying the effects of planar disorder on diffraction
spectra has a long history. (For a more complete discussion, see
Sebastian and Krishna.~\cite{sk94}) Perhaps the first quantitative
analysis was given by Landau~\cite{l37a} and Lifschitz~\cite{l37b}
assuming no correlation between MLs. Wilson~\cite{w42} provided an
analysis of planar imperfections in hexagonal Co by considering the
effect of the disorder on the Bragg peaks. This approach is necessarily
limited to the case of small amounts of faulting. Hendricks and
Teller~\cite{ht42} treated the problem rather generally, allowing for
different form factors for the different kinds of MLs, variable
spacing between MLs, and correlations between neighboring MLs. 
Since their method relies on extensive matrix calculations, it was found
cumbersome and difficult to apply by early researchers.

Jagodzinski developed a theory of diffraction for planar disorder by
considering nearest-neighbor correlations for general layered
structures~\cite{j49a} and for next-nearest neighbor correlations of
CPSs.~\cite{j49b} By noting the direction, length, and intensity of
non-Laue streaks in x-ray diffraction spectra of Cu-Si alloys,
Barrett~\cite{b50} was able to estimate the kind and approximate amount
of stacking disorder present.  Patterson~\cite{p52} considered the effect
of deformation faults on face-centered cubic crystals (fcc or 3C~\cite{note3}) and
demonstrated how one can calculate the fraction of faulted layers from
the widths and displacements of the Bragg peaks.  Gevers~\cite{g54b}
demonstrated how to calculate the effects of growth faults with an
$n$-layer influence on the diffraction spectra of close-packed crystals.
He also demonstrated how to treat both growth and deformation faults
randomly inserted into hexagonal (hcp or 2H) and cubic crystals, as well
as deformation faults randomly distributed into 4H and 6H
crystals.~\cite{g54a} Johnson~\cite{j63} examined the effects on the
diffraction pattern of random extrinsic faulting (insertion of a ML) in
the 3C structure and compared this to the effects of intrinsic faulting
(deletion of a ML) on the Bragg peaks in the limit of small fault
probabilities.  Prasad and Lele~\cite{pl70} considered the effects of
faulting on the diffraction spectra of 4H crystals containing up to nine
kinds of randomly distributed faults.  Pandey and Krishna~\cite{pk76}
treated the similar case of the 6H close-packed structure containing a
random distribution of $14$ distinct intrinsic faults.  Pandey and
Krishna~\cite{pk77} also derived an expression for the intensity of
diffracted radiation from a 2H crystal containing any amount of randomly
placed deformation and growth faults. By measuring the broadening of the
diffraction maxima of a SiC crystal they were able to determine the
amount of each kind of faulting. Michalski~\cite{m88} developed a general
theory for the random distribution of single stacking faults for an
arbitrary periodic structure and Michalski, {\it et al.}~\cite{mkd88} 
applied it to several hexagonal and rhombohedral
structures.

In an effort to understand experimental data concerning solid-state
transformations from the 2H to the 6H structure in annealed SiC crystals,
Pandey {\it et al.}~\cite{plk80a,plk80b,plk80c} developed the concept
of \emph{nonrandom faulting}.  They suggested that the presence of a
fault in a structure affects the probability of finding another fault
in close proximity.  They considered two possible faulting
mechanisms---deformation faulting and layer-displacement faults---and were
able to understand the stacking in SiC within this model. Similar work was
done on the 2H-to-3C transformation in ZnS
crystals.~\cite{s88,sk84,sk87a,sk87b,spk82,sk94,fjs86,plk80a,plk80b,plk80c,pl86a,pl86b,j72}

All of the above methods center on finding analytical expressions for
the diffracted intensity as a function of fault probabilities. In order
to make a quantitative estimate of the faulting, typically the placement,
broadening, shape, and symmetry of the Bragg peaks are compared with that
expected for a crystal containing a particular kind of fault.
Then the full width at half-maximum (FWHM) of one or several of the
Bragg peaks are used to determine the fraction of faulting.  

Thus, all of these approaches are limited to small amounts of disorder that
preserve the integrity of the Bragg peaks. (An exception to this is
Jagodzinski's~\cite{j49b} \emph{disorder model}. This approach has several
features in common with our own and can in fact be thought of as a
constrained case of our approach. We do not discuss his work further here,
but treat it and its relation to ours elsewhere.~\cite{vccup}) 
Should the disorder become
sufficiently large, the Bragg peaks become too broad and do not stand out
sufficiently from the diffuse, background scattering. We call these kinds
of approaches \emph{indirect}, because one begins with a set of postulated
faults and then sorts though them, searching for one or several that best
fit the data.

\subsection{Direct Methods}

Perhaps the first efforts at a \emph{direct} method for determining the structure
of disordered close-packed crystals were given by Dornberger-Schiff~\cite{d72}
and Farkas-Jahnke.~\cite{f73a,f73b} Dornberger-Schiff gave an algorithm
for relating Patterson values (i.e., fractions of faulted layers from Bragg
peaks) to sequence probabilities but, to our knowledge, did not follow up
with a method for finding the Patterson values from spectra showing diffuse
scattering. Farkas-Jahnke used Patterson-like functions to estimate the
frequency of occurrence of layer sequences up to length five. He was not
able to obtain a complete set of equations and this forced the use of
inequalities derived from symmetry arguments that do not generally hold in
disordered crystals. Unlike the previously considered techniques, these
methods are direct since they make no assumption about underlying crystal
structure or the faults it may contain. To our knowledge, since their
introduction three decades ago, neither of these methods have been used to
discover stacking structure in real materials. Hence we do not treat them
further here. 

\subsection{The Fault Model}

We refer to the indirect approaches---analyzing a crystal structure assuming it
contains a distribution of stacking errors or faults---as the \emph{fault model}
(FM). To date, it has been the dominant way in which planar disorder in
crystals has been viewed. However, we find a number of difficulties with the
FM, many of which have been recognized by previous
researchers.~\cite{g00,pw98,pp76a,f73a,tnd91,m88}
Our objections to the FM and the way it has been used to discover
structural information are severalfold.

\begin{trivlist}

\item ({\it i}) \emph{The FM assumes a parent crystal.} For the fault
model to make sense, it is necessary to assume some crystal structure
in which to introduce faulting. This may be satisfactory for weakly
faulted crystals, but for those with significant disorder or those
undergoing a solid-state phase transition to another crystal
structure,~\cite{fjs86,j72,kp95,km71a,km71b,pl86a,pl86b,plk80a,plk80b,plk80c,spk82,sk87a,sk87b,sk87c,snk87,sp96,
sp96a,stkp96,sp97} this picture is untenable.

\item ({\it ii}) \emph{Each parent crystal must be treated separately.}
Since the FM introduces stacking ``mistakes'' into a parent crystal, each
kind of parent crystal must be treated individually. There has been
significant work on only two CPSs; namely, the 2H and 3C. In polytypism
hundreds of other crystalline structures are known to exist. In the FM, each
must be analyzed separately by postulating appropriate crystal-specific kinds
of defects. Given this degree of complication, it is desirable to find a
theoretical framework that unites the description of the various kinds of fault
and parent structure into a single, coherent picture.

\item ({\it iii}) \emph{In practice, the FM treats only the Bragg peaks
quantitatively, effectively ignoring the diffuse scattering.} Most
researchers use formul\ae\ that give the FWHM of Bragg peaks in terms
of the fraction of certain postulated defects to find the amount of
faulting. Some do acknowledge that diffuse scattering is important, but
to our knowledge none use the diffuse scattering to \emph{quantitatively}
measure crystal structure.~\cite{note6}

\item ({\it iv}) \emph{The FM is unable to capture the variety of naturally
occurring stacking sequences.} By assuming a small set of possible ways
that a parent structure can deviate from crystallinity, the FM necessarily
assumes that there are stacking sequences that do \emph{not} occur.
It is
desirable to have an approach that considers as many candidate structures
as possible with as few \emph{a priori} restrictions as possible.

\item ({\it v}) \emph{The FM's description of the disorder is not unique.}
It is possible to give two different faulting schemes that describe the
same weakly faulted material.~\cite{vcc02a} This is readily seen by noting
that the layer-displacement fault in a 2H crystal can be viewed as two adjacent, but
oppositely oriented deformation faults.

\end{trivlist}

\subsection{Computational Mechanics}

Our method of discovery and quantification of planar structure and disorder
in crystals
overcomes all of these difficulties for the special but important case
of CPSs. We do not assume an underlying crystalline structure. Indeed, we
make no assumptions at all about either the crystal or fault structure
that may be present. Instead, we find the frequency of occurrence of
all possible stacking sequences up to a given length and use this to
construct a model that captures the statistics of the stacking sequence.
In this sense, we directly determine the stacking structure.  Our scheme
for describing planar disorder unites both fault and crystal structure
into a single framework. There is no need to treat each crystal structure
or faulting scheme separately. Our method treats any amount and kind of
planar disorder present.  Finally, we quantitatively use all of the
information contained in the diffraction spectra, both in Bragg peaks
and in diffuse scattering, to build a unique model of the stacking structure.
This model does \emph{not} find the particular stacking sequence of the
specimen that generated
the diffraction pattern---this is not possible from diffraction spectra
alone---but rather finds the statistical regularities across an ensemble
of stacking sequences that could have given rise to the observed spectra.
This is the best that can be done, in principle.

The history of discovering planar disorder in crystals is one of consistent,
incremental progress over nearly seventy years. There are two factors,
however, that have hindered progress in this area. The first is
calculational. Much of the early work centered on finding analytical
expressions for the diffracted intensity of a given crystalline structure
permeated with a particular fault type. With the advent of modern
numerical and symbolic calculational methods and the concomitant ability
to estimate diffraction patterns from any arbitrarily layered
structure,~\cite{tnd91} much of this early work has been superseded. The
second hindrance to progress has been a lack of fundamental understanding of
structure and disorder in one-dimensional sequences generated by nonlinear
dynamical systems. Recently, however, a unifying framework has been introduced
in theory of \emph{computational
mechanics}.~\cite{cy89,cf97,cf01,fc98a,sc01,f98a}

Computational mechanics is an approach to discovering, describing, and
quantifying patterns. It provides for the construction of the minimal and
unique model for a process that is optimally predictive; this model is
called an \emph{\eM}. A process's \eM\ is minimal in the sense of requiring
the fewest model components to represent the process's structures and disorder;
it is optimal in the sense that no alternative representation is more accurate;
and it is unique in the sense that any alternative which is both minimal and
optimally predictive is isomorphic to it. An \eM's algebraic structure
captures a process's symmetries and approximate symmetries. From an \eM\ 
measures of a process's memory, entropy production, and structural complexity
can be found. We demonstrate in the sequel~\cite{vcc02c} that knowledge of
the \eM\ and the energy coupling between MLs allows one to calculate the
average stacking energy for a disordered polytype. Before being adapted to
the present application to polytypes, computational mechanics had been
used to analyze structural complexity in a wide range of nonlinear processes,
such as cellular automata,~\cite{h93} the logistic map,~\cite{cy89,cy90}
and the one-dimensional Ising model,~\cite{f98a,cf97} as well to experimental 
physical systems, such as the dripping faucet,~\cite{gpso98} atmospheric
turbulence,~\cite{pfb00} and geomagnetic data.~\cite{cfw01}

Our development here is organized as follows: in \S
\ref{SpectralEMReconstruction} we give a detailed account of our procedure for
discovering and quantifying disorder in CPSs; in \S
\ref{EMFaultModelStructuralAnalyses} we compare our approach to the FM; in
\S \ref{ExampleProcesses} we treat four prototype polytypes to demonstrate
our technique; in \S \ref{CPSCharacteristicLengths} we discuss several
characteristic lengths that can be estimated from our model and we address
how they relate to the range of interactions between MLs; and in \S
\ref{Conclusions} we give our conclusions.

\section{\EM\ Spectral Reconstruction}
\label{SpectralEMReconstruction}

Previous techniques of \eM\ reconstruction have used a sequence of data
produced by the process.~\cite{cy89,c94,Shal02a} Here, the experimental signal
comes in the form of a power spectrum, and we need to develop a technique to 
infer the \eM\ from this type of data. We call this new class of inference
algorithms \emph{\eM\ spectral reconstruction}---abbreviated \eMSR\ and
pronounced ``emissary''. We emphasize that our goal remains unchanged---to
find the process's underlying description. It is only the inference procedure
that is changed. In this section we give a detailed account of \eMSR\ as
applied to the problem of discovering pattern and disorder in CPSs. 
 
We divide \eMSR\ into five steps. First, we extract correlation
information from a diffraction spectrum. Second, we use this to estimate
stacking-sequence probabilities of a given length. Third, we reconstruct
an \eM\ from this distribution. Fourth, we generate a diffraction spectrum
from the \eM. And, finally, we compare this \eM\ spectrum to the original.
If there is insufficient agreement, we repeat the second through fourth
steps, estimating stacking-sequence probabilities at a longer length,
building a new \eM, and again comparing with the original spectrum. In the
final two subsections, we give relations that can be used to determine
the quality of experimental data and briefly review several
information- and computation-theoretic quantities of physical
import that can be directly estimated from the reconstructed \eM.

\subsection{Correlation Factors from Diffraction Spectra}
\label{CorrelationFactorsfromDiffractionSpectra}

We start with the conventional assumptions concerning polytypism in CPSs.
Namely, we assume that
\begin{itemize}
\setlength{\topsep}{0mm}
\setlength{\itemsep}{0mm}
\item the MLs themselves are undefected and free of any distortions;
\item the spacing between MLs does not depend on the local stacking arrangement;
\item each ML has the same scattering power; and
\item the faults extend completely across the crystal.
\end{itemize}
We make the additional assumption that the probability of finding a given
stacking sequence in the crystal remains constant through the crystal. (In
statistics parlance, we assume that the process is \emph{stationary}.)

Let $N$ be the number of hexagonal, close-packed MLs, with each ML occupying
one of three orientations, denoted $A$, $B$, or $C$.~\cite{am76,k96,sk94,vc01}
We introduce three statistical quantities, $Q_c(n)$, $Q_a(n)$, and
$Q_s(n)$:\cite{yc96} the two-layer \emph{correlation functions} (CFs),
where $c$, $a$, and $s$ stand for \emph{cyclic}, \emph{anti-cyclic}, and
\emph{same}, respectively. $Q_c(n)$ is defined as the probability that any
two MLs at a separation of $n$ are cyclically related. By cyclic, we mean
that if the  $i^{\mathrm {th}}$ ML is in orientation $A$ ($B, C$), say, then the
$(i+n)^{\mathrm {th}}$ ML is in orientation $B$ ($C, A$). $Q_a(n)$ and $Q_s(n)$ are
defined in a similar fashion. Since these are
probabilities, $0 \leq Q_{\alpha}(n) \leq 1$, where $\alpha \in \{c,a,s\}$.
Additionally, at each $n$ it is clear that $\sum_{\alpha} Q_{\alpha}(n) = 1$.

With these assumptions and definitions in place, the \emph{total diffracted
intensity} along the $10.l$ row~\cite{note4} can be written as~\cite{g63,bw86,yc96}
\begin{eqnarray}
  I(l) & = & \psi^2 \biggl( \frac{\sin^{2}(\pi Nl)}{\sin^{2}(\pi l)}
  -2\sqrt{3}\sum_{n=1}^{N}\Bigl\{(N-n) \times  \nonumber \\
       &   & \mbox{} \bigl[Q_c(n)\cos(2\pi nl+\frac{\pi}{6})  \nonumber \\
	   &   & \mbox{} + Q_a(n) \cos(2 \pi nl - \frac{\pi}{6})\bigr]
       \Bigr\} \biggr),
\label{eq:diff1}
\end{eqnarray}
where $l$ is a continuous variable that indexes the magnitude
of the perpendicular component of the diffracted wave, $k = 2\pi l/c$,
and $c$ is the spacing between adjacent MLs.  $\psi^2$ is a function of
$l$ that accounts for atomic scattering factors, the structure factor,
dispersion factors, or any other effects for which the experimentally
obtained diffraction spectra may need to be corrected.~\cite{hws92,w97,m73}

It is convenient to work with the intensity per ML, instead of the total
intensity, so we define the corrected diffracted intensity per ML, $\mathsf{I}(l)$, as
\begin{eqnarray}
   \mathsf{I}(l) = \frac{I(l)}{\psi^2N} ~.
\label{eq:diff2}
\end{eqnarray}
We will always use $\mathsf{I}(l)$ unless otherwise noted and simply call
this the \emph{diffracted intensity}. Observe that the diffracted intensity
$\mathsf{I}(l)$ integrated over any unit $l$-interval is unity regardless of
the particular values of the CFs.~\cite{v01} We may then use this fact to
normalize experimental data.

The form of Eqs. (\ref{eq:diff1}) and (\ref{eq:diff2}) suggests that
the CFs can be found from the diffraction pattern by Fourier analysis.
Let us define $\mathsf{X}(n)$ and $\mathsf{Y}(n)$ as
\begin{eqnarray}
   \mathsf{X}(n) = \oint \mathsf{I}(l) \cos (2 \pi nl) \, dl
\label{eq:def_Xn}
\end{eqnarray}
and
\begin{eqnarray}
   \mathsf{Y}(n) = \oint \mathsf{I}(l) \sin (2 \pi nl) \, dl ~,
\label{eq:def_Yn}
\end{eqnarray}
where the small circle in the integral sign indicates that
the integral is to be taken over a unit interval in $l$. It is possible
to show~\cite{v01} that in the limit $N \rightarrow \infty$
\begin{eqnarray}
   Q_c(n) = \frac{1}{3} - \frac{1}{3} \Bigl[  \mathsf{X}(n) -
   \sqrt{3}\mathsf{Y}(n) \Bigr]
\label{eq:def_Qcn}
\end{eqnarray}
and
\begin{eqnarray}
   Q_a(n) = \frac{1}{3} -  \frac{1}{3} \Bigl[ \mathsf{X}(n) +
   \sqrt{3}\mathsf{Y}(n) \Bigr] ~.
\label{eq:def_Qan}
\end{eqnarray}
Thus, the CFs can be found by Fourier analysis of the diffraction spectrum.

\subsection{Estimating the Stacking-Sequence Distribution}
\label{EstimateStackingSequenceDistribution}

In the second part of our approach, we estimate the distribution of stacking
sequences from the two-layer CFs.
First, though, we must consider what kind of information the CFs contain
about stacking sequences. Therefore let us define the \emph{stacking process}
as the effective stochastic process induced by scanning the stacking sequence
along the stacking direction. It is convenient to represent the stacking
sequence in terms of the H\"{a}gg notation,~\cite{sk94} where one replaces
the set of allowed orientations $\{A,B,C\}$ of a ML with a binary alphabet
$\mathcal{A} = \{0,1\}$. On moving from the $i^{\mathrm {th}}$ to the
$(i+1)^{\mathrm {th}}$ ML, we label each inter-ML transition or
\emph{spin}~\cite{vc01} as ``1'' if the two MLs are cyclically related ($A
\rightarrow B \rightarrow C \rightarrow A$) and ``0'' if the two MLs 
are anti-cyclically related ($A \rightarrow C \rightarrow B \rightarrow
A$). Thus, the stacking constraint that no two adjacent MLs may have
the same orientation ($A$, $B$, or $C$) is built into the notation. There
is a one-to-one mapping between the stacking orientation sequence and the
spin sequence, up to an overall rotation of the crystal; and we use them
interchangeably.

We estimate the probability distribution $\Prob(\omega)$ of finding sequences
$\omega$ averaged over the sample by considering a series of constraints
on the sequence probabilities. Some of these constraints are simple
consequences of the mathematics; some come from the CFs themselves.
From conservation of probability, we have
\begin{eqnarray}
   \Prob(u) = \Prob(0u) + \Prob(1u) = \Prob(u0) + \Prob(u1) ~,
\label{eq:cons.of prob.1}
\end{eqnarray}
for all $u \in \ProcessAlphabet^\Range$, where $\ProcessAlphabet^{\Range}$
is the set of all sequences of length $\Range$ in the H\"{a}gg notation.
Additionally, we require that the sum of all probabilities of sequences
of length $\Range + 1$ be normalized, {\it i.e.},
\begin{eqnarray}
   \sum_{\omega \in \mathcal{A}^{\Range+1}} \Prob(\omega) = 1 ~.
\label{eq:cons.of prob.2}
\end{eqnarray}
Equations~(\ref{eq:cons.of prob.1}) and (\ref{eq:cons.of prob.2})
together provide $2^{\Range}$ constraints among the $2^{\Range+1}$
possible stacking sequences of length $\Range + 1$.

The remaining $2^\Range$ constraints come from relating CFs to sequence
probabilities via the relations
\begin{eqnarray}
   Q_{\alpha}(n) = \sum_{\omega \in \mathcal{A}_{\alpha}^n} \Prob(\omega) ~,
\label{eq:Q.to.p}
\end{eqnarray}
where $\mathcal{A}_{\alpha}^n$ is the subset of length-$n$ sequences that
generate a cyclic ($\alpha = c$) or an anti-cyclic ($\alpha = a$) rotation
between MLs at separation $n$. A sequence generates a cyclic (anti-cyclic)
rotation between MLs at separation $n$ if $2m-n = 1  \pmod 3$, where $m$
is the number of $1$s ($0$s) in the sequence.  We take as many of the relations
in Eq. (\ref{eq:Q.to.p}) as necessary to form a complete set of equations
to solve for $\Prob(\omega)$. At fixed $\Range$, the set of equations describes
the stacking sequence as an $\Range^{\mathrm {th}}$-order Markov process.
For $\Range = 1$ and $\Range = 2$ the sets of equations are linear and admit
analytical solutions. 
At $\Range = 3$, the first nonlinearities appear due to the necessity of using
CFs at $n=5$ to obtain a complete set of equations. We rewrite the conditional
probabilities at $n=5$ in terms of those at $n=4$ via relations of the form
\begin{eqnarray}
\Prob(s_0s_1s_2s_3s_4) & = & \Prob(s_0s_1s_2s_3)\Prob(s_4|s_0s_1s_2s_3)   \nonumber \\
    & \approx & \Prob(s_0s_1s_2s_3)\Prob(s_4|s_1s_2s_3) \nonumber \\
    & = & \frac{\Prob(s_0s_1s_2s_3)\Prob(s_1s_2s_3s_4)}{\Prob(s_1s_2s_30)+\Prob(s_1s_2s_31)} ~,
\label{eq:p.reduction}
\end{eqnarray}
where, in the second line, the approximation is invoked. We refer to this
approximation as \emph{memory-length reduction}, as it effectively limits
the memory that we consider in order to obtain a complete set of equations.  

We refer collectively to the set of Eqs. (\ref{eq:cons.of prob.1}),
(\ref{eq:cons.of prob.2}), and (\ref{eq:Q.to.p}) as the
{\em spectral equations at a given $\Range$.} The analytical solutions for
the $\Range = 1$ and $\Range = 2$ spectral equations are given in Appendix A, along
with the spectral equations for $\Range = 3$. For the $\Range = 3$ spectral
equations, we solve these numerically~\cite{v01} to find $\Prob(\omega)$. 

\subsection{\EM\ Reconstruction from the Stacking Process}
\label{EMFromStackingProcess}

In the third part of our approach, we infer the stacking process's
\eM\ from the estimated distribution of stacking sequences.

Suppose we know the probability distribution $\Prob(\omega)$ of stacking
sequences $\omega = \ldots s_{-2} s_{-1} s_0 s_1 s_2 \ldots$, where
$s_i \in \ProcessAlphabet$ and $\omega$ is a stacking sequence in the
H\"{a}gg notation. Then at each ML we define the ``past''
$\stackrel{\leftarrow}{\omega}$ as all the previous transitions $s_i$ seen
and the ``future'' $\stackrel{\rightarrow}{\omega}$ as those transitions
$s_i$ yet to be seen: that is, $\omega = \stackrel{\leftarrow}{\omega}
\stackrel{\rightarrow}{\omega}$.

The effective states or \emph{causal states} (CSs) of the stacking process are
defined as the \emph{sets} of pasts $\stackrel{\leftarrow}{\omega}$ that lead
to statistically equivalent futures:
\begin{eqnarray}
   \stackrel{\leftarrow}{\omega}_i ~\sim~ \stackrel{\leftarrow}{\omega}_j
   \, \textrm{ if and only if } \,
	 \Prob(\stackrel{\rightarrow}{\omega}|\stackrel{\leftarrow}{\omega}_i)
	 = \Prob(\stackrel{\rightarrow}{\omega}|\stackrel{\leftarrow}{\omega}_j) ~,
\label{eq:cm}
\end{eqnarray}
for all futures $\stackrel{\rightarrow}{\omega}$, where
$\Prob(\stackrel{\rightarrow}{\omega}|\stackrel{\leftarrow}{\omega}_i)$ is
the conditional probability of seeing $\stackrel{\rightarrow}{\omega}$
having just seen $\stackrel{\leftarrow}{\omega}_i$.~\cite{cy89,c94,sc01}

As a default set of CSs, we initially assume that each history of length
$\Range$ forms a unique CS. So, for \eMSR\ at $\Range$, we begin with
$2^{\Range}$ CSs, each labeled by its unique length-$\Range$ history. We
refer to this set of CSs as {\em candidate causal states}, as they may not
be the true CSs that describe the stacking process. We now estimate the
state-to-state transition probabilities between candidate CSs as follows.
Define the \emph{transition matrices}
${\sf T}_{i \rightarrow j}^{(s)}$ as the probability of making a transition
from a candidate CS $\CausalState_i$ to a candidate CS $\CausalState_j$ on
seeing spin $s$. If we label each past by the last $\Range$ spins seen,
then this implies that only transitions of the form $s_0v \rightarrow vs$ are 
allowed, where $v \in \mathcal{A}^{r-1}$. All other transitions are taken to be zero.    
Then we can write the transition matrix as
\begin{eqnarray}
   {\sf T}_{i \rightarrow j}^{(s)} = {\sf T}_{s_0v \rightarrow vs}^{(s)}.
\label{eq:em_transitions_def}
\end{eqnarray}
We estimate these transition probabilities from the conditional probabilities,
\begin{eqnarray}
   {\sf T}_{s_0v \rightarrow vs}^{(s)} & \approx & \Prob(s|s_0v)  \nonumber  \\
                                     & = & \frac{\Prob(s_0vs)}{\Prob(s_0v)}.
\label{eq:em_transitions_est}
\end{eqnarray}

We now apply the the equivalence relation, Eq. (\ref{eq:cm}), to merge
histories with equivalent futures. The set of resulting CSs, along with
the transitions between states, defines the process's \emph{\eM}. This is
the minimal, unique description of the stacking process that optimally
produces the stacking distribution $\Prob(\omega)$. At this point,
we should refer to this as a {\em candidate} \eM, 
as it will reproduce the CFs used to find it, but it may fail to reproduce
CFs at larger $n$ satisfactorily. The next two subsections address this
issue of agreement between theory and experiment.  

It is worth repeating that our method of \eM\ reconstruction is novel
in the sense that we do not estimate sequence probabilities from a long
string of symbols generated by the process, as has been done
previously.~\cite{c94,Shal02a} Rather we use the two-layer CFs obtained
from Fourier analysis of the diffraction spectra. In this way, 
\eMSR\ is accomplished purely from spectral information.

\subsection{\EM\ Correlation Factors and Spectrum}

In the fourth part, we use the reconstructed \eM\ to generate a sample
spin sequence $M$ spins long in the H{\"{a}}gg representation. We then change
representations by mapping this spin sequence to a stacking-orientation
sequence in the ($A,B,C$) notation. We directly find the CFs by scanning the
stacking-orientation sequence. The CFs for all of the processes we consider
decay to an asymptotic value of $1/3$ for large $n$. We set the CFs to $1/3$
when they reach $\approx 1\%$ of this value, which occurs typically for
$n \approx 25-100$. We could, of course, find the CFs directly from
spin-sequence probabilities, via Eq. (\ref{eq:Q.to.p}). However, if one needed
to calculate CFs for, say, $n=50$, this would require finding the sequence
probabilities for sequences of length $50$. There are $2^{50} \approx 10^{15}$
spin sequences for $n=50$, so the sum implied by Eq. (\ref{eq:Q.to.p}) is
difficult to perform in practice. As an alternative, one changes representation
and rewrites the candidate \eM\ in terms of the absolute stacking positions,
$\{A, B, C\}$. From this new representation the CFs are calculable from the
transition matrices. Additionally, it is possible to derive analytical
expressions for the CFs in some cases.~\cite{v01} This has not been done
here, however.
 
Once the CFs have been found, the diffraction spectrum is readily calculated
from Eqs. (\ref{eq:diff1}) and (\ref{eq:diff2}). It has been shown that for
sufficiently large $M$, the diffraction spectrum for diffuse scattering scales
as $M$,~\cite{v01} so that the number of MLs used to calculate the diffraction
spectrum is not important, if $M$ is sufficiently large (say, 10 000). To
reduce the error due to fluctuations, it is desirable use as long a
sequence as possible to find the CFs.  

In this way, the \eM's predicted CFs and diffraction spectrum can be
calculated.  

\subsection{Comparing Original with \EM\ Spectra}

In the fifth and final part, we compare \eM\ CFs and spectrum to the
original spectrum. If there is not sufficient agreement, we increment
$\Range$ and repeat the reconstruction and comparison.

More precisely, in comparing the reconstructed \eM\ ``theory'' with the
original spectra, we need a quantitative measure of the goodness-of-fit
between them. We use the \emph{profile $\cal R$-factor},~\cite{note5} which
is defined as
\begin{equation}
{\cal R} =
 \frac{\oint \bigl|\mathsf{I}_{\epsilon {\rm M}}(l) - \mathsf{I}_{exp}(l)\bigr| \,dl}
 {\oint \mathsf{I}_{\epsilon {\rm M}} \, dl} \times 100 \% ~,
\label{R_def}
\end{equation}
where $\mathsf{I}_{\epsilon {\rm M}}(l)$ is the \eM\ diffraction spectrum and
$\mathsf{I}_{exp}(l)$ is the experimental. Notice that the denominator is
unity due to normalization.

It is important, however, not to over-fit the original data, so we should not
seek a fit that is closer than experimental error. Let us define
$\delta \mathsf{I}_{exp}(l)$ as the fluctuation-induced error in the diffracted
intensity as a function of $l$. Then \emph{spectral error} $R_{err}$ can be
defined as
\begin{eqnarray}
   {\cal R}_{err} = \frac{\oint \bigl|\delta \mathsf{I}_{exp}(l) \bigr| \, dl } 
   {\oint \mathsf{I}_{exp} \, dl} \times 100 \% ~.
\label{Rerr_def}
\end{eqnarray}
Notice that the denominator once again reduces to unity due to normalization.
${\cal R}_{err}$ gives a measure of how two diffraction spectra taken from the
same sample over the same interval will differ from each other. Clearly, we
do not wish to seek an \eM\ that gives better agreement than this. So our
criteria for stopping reconstruction is when
$|{\cal R} - {\cal R}_{err}| \leq \Gamma$, where the acceptable-error threshold
$\Gamma$ is set in advance.

\subsection{Figures-of-Merit for Spectral Data}
\label{SpectralFiguresofMerit}

An issue we have so far neglected is the CFs' independence. In order to solve
the spectral equations, part 3 in \eMSR\ (\S \ref{EMFromStackingProcess}), 
we need $2^{r+1}$ independent constraints. It is therefore important to
identify and avoid using any redundancies inherent in the CFs to solve the
spectral equations. Rather than finding this a hindrance, any relations 
that CFs obey can be exploited to assess the quality of experimental data
over a given $l$-interval. We find that, as a result of stacking constraints
and conservation of probability, there are two equalities that the CFs must
satisfy. We develop and define these measures here. 

We find the first by observing that, at $n=1$, due to stacking constraints,
$Q_c(1) + Q_a(1) = 1$. Adding Eqs.~(\ref{eq:def_Qcn}) and (\ref{eq:def_Qan})
with $n=1$ immediately gives $\mathsf{X}(1) = -1/2$. This suggests that we
define a \emph{figure-of-merit} $\gamma$ as
\begin{eqnarray}
     \gamma = \oint \mathsf{I}(l) \cos (2 \pi l) \, dl ~.
\label{eq:x1}
\end{eqnarray}
$\gamma$ can be used to evaluate the quality of experimental spectra. For
an ideal, error-free spectrum, $\gamma = -1/2$. Since many spectra
are known to contain some systematic error,~\cite{pplg87,sk94}
the amount by which $\gamma$ deviates from $-1/2$ can be used
to assess how corrupt the data is over a given unit $l$-interval.

To find the second constraint, we observe that Eq.~(\ref{eq:cons.of prob.1}),
with $\Range=1$ and $u=0$, gives $\Prob(01) = \Prob(10)$. We therefore find from
Eq.~(\ref{eq:cons.of prob.2}) that $\Prob(00) + 2\Prob(01) + \Prob(11) = 1$.  We can
write $\Prob(01) = \Prob(1) - \Prob(11)$. This implies that
\begin{equation}
\Prob(00) + 2\Prob(1) -\Prob(11) = 1 ~.
\end{equation}
Making the identification from Eq.~(\ref{eq:Q.to.p}) that $Q_c(1) = \Prob(1)$,
$Q_a(2) = \Prob(11)$, and $Q_c(2) = \Prob(00)$ gives
\begin{equation}
2Q_c(1) + Q_c(2) - Q_a(2) = 1 ~.
\end{equation}
This suggests that we define a second \emph{figure-of-merit} $\beta$ to be
\begin{eqnarray}
    \beta = 2Q_c(1) + Q_c(2) - Q_a(2) ~.
\label{eq:x2}
\end{eqnarray}
$\beta$ should be unity for error-free data. This can also be used to evaluate
the quality of the experimental data over a given unit $l$-interval. Together,
$\gamma$ and $\beta$ are the figures-of-merit over a unit $l$-interval for a
diffraction spectrum. Therefore, in the first part of \eMSR\
(\S \ref{CorrelationFactorsfromDiffractionSpectra}) we evaluate 
each over candidate $l$-intervals and choose an interval for
\eM\ reconstruction that gives figures-of-merit best in agreement with the
theoretical values. These two constraints on the CFs imply that only
two out of the first four correlation functions, $Q_c(1)$, $Q_a(1)$,
$Q_c(2)$, and $Q_a(2)$ are independent. We choose to take the $n=2$
terms as the independent parameters in the spectral equations.  

This completes our presentation of \eM\ spectral reconstruction.
The overall procedure is summarized in Table \ref{SpectraleMReconstruction}.

\begin{table}
\begin{center}
\begin{tabular}{l}
\hline
\hline
1. Find the CFs from the diffraction spectrum.\\
~~~~1a. Correct the spectrum for any experimental factors. \\
~~~~1b. Calculate the figures-of-merit (\S \ref{SpectralFiguresofMerit})
over possible \\
~~~~~~~~~$l$-intervals to find an interval suitable for \eM\ \\
~~~~~~~~~reconstruction.\\
~~~~1c. Find the CFs over this interval.\\
~~~~1d. Estimate the spectral error ${\cal R}_{err}$ from the \\
~~~~~~~~diffraction spectrum.\\
2. Estimate stacking distribution $\Prob(\omega^\Range)$ from CFs. \\
~~~~2a. Set $\Range = 1$. \\
~~~~2b. Solve the spectral equations for $\Prob(\omega^\Range)$. \\
3. Reconstruct the \eM\ from the $\Prob(\omega^\Range)$. \\
~~~~3a. Label candidate CSs by their length-$\Range$ histories. \\
~~~~3b. Estimate transition probabilities between states \\
~~~~~~~~~from sequence probabilities.  \\
~~~~3c. Merge histories with equivalent futures to form \\
~~~~~~~~~CSs. \\
4. Generate CFs and diffraction spectrum from the \\
~~~~\eM. \\
5. Calculate the error $\Gamma(\Range) = |{\cal R} - {\cal R}_{err}|$ between\\
~~~~the original and \eM\ spectra:\\
~~~~5a. If $\Gamma(\Range) \geq \Gamma$, replace $\Range$ with $\Range + 1$
        and go to step 2b;\\
~~~~5b. Otherwise, stop.\\
\hline
\hline
\end{tabular}
\caption{The \eMSR\ algorithm. Here $\omega^\Range$ signifies the set of
  length-$\Range$ sequences.
  }
\label{SpectraleMReconstruction}
\end{center}
\end{table}

\subsection{Measures of Structure and Intrinsic Computation}
\label{MeasuresStructureIntrinsicComputation}

There are a number of different quantities in computational mechanics
that describe the way information is processed and stored. (See Crutchfield
and Feldman~\cite{cf01} and Shalizi and Crutchfield~\cite{sc01} for a detailed
discussion and mathematical definitions.) We consider only the following.
\begin{trivlist}
\item \emph{Memory Length} $\MLength$: The value of $\Range$ that results at
	the termination of \eMSR\ is an estimate of the
	stacking process's \emph{memory length}, denoted $\MLength$, since it is
	the number of MLs that one must use to optimally represent the process's
	sequence statistics (given the accuracy of the original spectrum).
\item \emph{Statistical Complexity} $\Cmu$: The minimum average amount of
	memory needed to statistically reproduce a process is known as the
	statistical complexity $\Cmu$. Since this a measure of memory, it has
	units of [\emph{bits}]. It is the Shannon information stored in the
	set of CSs: 
        \begin{equation}
        	\Cmu = - \sum_{\causalstate \in \CausalStateSet}
        	\Prob(\causalstate) \log_2 \Prob(\causalstate) ~,
        \end{equation}
	where $\CausalStateSet$ is the set of CSs for the process and 
	$\Prob(\causalstate)$ is the asymptotic probability of CS $\causalstate$.
	The latter is the left eigenvector, normalized in probability, of the
	state-to-state transition matrix
	${\sf T} = \sum_{s \in \ProcessAlphabet} {\sf T}^{(s)}$. Physically, the
	statistical complexity is related to the \emph{average} number of previous
	spins one needs to observe on scanning the spin sequence to make an optimal
	prediction of the next spin. The statistical complexity is also related to
	a generalization of the stacking period for nonperiodic processes. We
	detail this connection in \S \ref{CPSCharacteristicLengths}. 
\item \emph{Entropy Rate} $\hmu$: The amount of irreducible randomness
	per ML after all correlations have been accounted for. It has units
	of [\emph{bits/ML}]. It is also known as the \emph{thermodynamic entropy
	density} in statistical mechanics and the \emph{metric entropy} in
	dynamical systems theory. It is given by the average per-state uncertainty:
\begin{equation}
	\hmu = - \sum_{\causalstate \in \CausalStateSet} \Prob(\causalstate)
	\sum_{s \in \ProcessAlphabet}
	{\sf T}_{\causalstate \rightarrow \causalstate^\prime}^{(s)}
	\log_2 {\sf T}_{\causalstate \rightarrow \causalstate^\prime}^{(s)}
	~,
\end{equation}
where $\causalstate^\prime$ is the CS reached from $\causalstate$
upon seeing spin $s$. Physically, $\hmu$ is a measure of the entropy density
associated with the stacking process.    
\item \emph{Excess Entropy} $\EE$: The amount of \emph{apparent} memory in a
	process. The units of $\EE$ are [\emph{bits}]. It is defined as the amount
	of Shannon	information shared between the left and right halves of a
	stacking sequence:
\begin{equation}
\EE = \sum_\omega \Prob(\omega) \log_2
	\frac{\Prob(\omega)}
	{\Prob(\stackrel{\leftarrow}{\omega})\Prob(\stackrel{\rightarrow}{\omega})}
\end{equation}
\end{trivlist}

Note that Crutchfield and Feldman~\cite{fc98a,cf01} showed that, for
range-$\Range$ Markov processes, these quantities are related by
\begin{equation}
\Cmu = \EE + \Range \hmu ~.
\label{BlockMarkovComplexityReln}
\end{equation}
For general nonfinite-range Markov processes, at present all that can be said
is that the statistical complexity upper bounds the excess entropy:
$\EE \leq \Cmu$.~\cite{sc01}

\section{\EM\ and Fault-Model Structural Analyses}
\label{EMFaultModelStructuralAnalyses}

\begin{figure}
\begin{center}
\resizebox{!}{1.75in}{\includegraphics{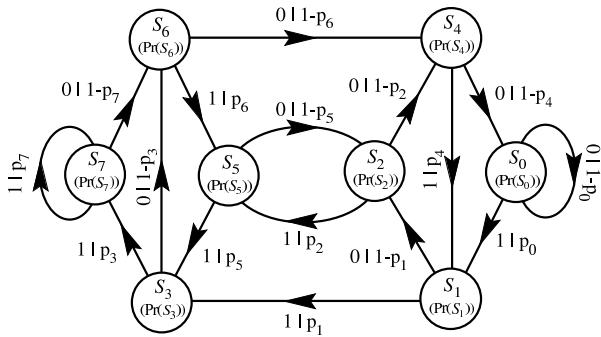}}
\end{center}
\caption{The most general $\Range=3$ \eM. We show only the recurrent portion of
  the \eM, as the transient part is not physically relevant (at this stage).
  The CSs are labeled by the last three spins seen, {\it i.e.} {\bf {$S_5$}}
  means that 101 were the last three spins seen. The numbers in parentheses
  are the asymptotic CS probabilities. The edge label $s|p$ indicates a
  transition on spin $s$ with probability $p$.
  }
\label{fig:dB3.general}
\end{figure}

Now that a statistical description of the stacking process has been found in the
form of an \eM, it is desirable to give an intuitive notion of what the
structure of the \eM\ tells us about the patterns and disorder in a stacking
process. In this section, we define and discuss architectural features of
$\Range = 3$ \eMs\ and their relation to the FM. Specifically, we discuss
the form that growth, deformation, and layer-displacement faulting for the
2H and 3C structures assume on an $\Range = 3$ \eM. With this connection in
place, we then address the general interpretation of \eMs\ as related to the
stacking of CPSs and demonstrate the superiority of the \eM\ as a general
indicator of structure and disorder in CPSs. 

\subsection{Causal-State Cycles}

\subsubsection{Definitions}

Since the \eM\ reconstructed at $\Range$ can distinguish at most only
$2^{\Range}$ pasts, it can have no more than $2^{\Range}$ CSs. The most
general reconstructed \eM\ of memory length $\Range$ is topologically
equivalent to a \emph{de Bruijn graph}~\cite{t90} of order $\Range$.
By ``most general'' we mean that all length-$\Range$ pasts are distinguished
and all allowed transitions between CSs exist. Under these assumptions,
the most general binary $\Range=3$ \eM\ (which has $2^3 =8 $ CSs and
$2^{3+1} = 16$ transitions) is shown in Fig.~\ref{fig:dB3.general}.
It is known that de Bruijn graphs can broken into a finite number of
closed, nonself-intersecting loops called \emph{simple cycles}
(SCs).~\cite{cw96}

By analogy, we define a \emph{causal-state cycle} (CSC) as a finite, closed,
nonself-intersecting, symbol-specific path on an \eM. We denote a CSC by
the sequence of CSs visited in square brackets \mbox{[  ]}. The states
themselves are labeled with a number (in decimal notation) that gives the
sequence of the last three spins leading to that CS. For example, for an
$\Range = 3$ reconstructed \eM, CS $\CausalState_3$ means that $011$ were
the last three spins observed before reaching that CS. The \emph{period} of
the CSC is the number of CSs that comprise it.

We further divide CSCs into \emph{strong} and \emph{weak} depending on the
strengths of the transitions between the CSs that make up the CSC.
The \emph{causal-state cycle probability} $\CSCP$ is defined as the cumulative
probability to complete one loop of a CSC, beginning on the CSC. We identify
CSCs with large $\CSCP$ as strong CSCs and all others as weak CSCs.

\begin{table}
\begin{center}
\begin{tabular}{l|l|l}
\hline
\hline
[$\CausalState_0$]     & (0)$^*$         & 3C    \\ \hline
[$\CausalState_7$]     & (1)$^*$         & 3C    \\ \hline 
[$\CausalState_2\CausalState_5$]     & (01)$^*$        & 2H              \\ \hline
[$\CausalState_1\CausalState_3\CausalState_6\CausalState_4$]     & (0011)$^*$      & 4H  \\ \hline
[$\CausalState_1\CausalState_3\CausalState_7\CausalState_6\CausalState_4\CausalState_0$]     & (000111)$^*$    & 6H \\ \hline
[$\CausalState_5\CausalState_2\CausalState_4\CausalState_1\CausalState_3\CausalState_7$] & (001101)$^*$    & 6H$_a$  \\ \hline
[$\CausalState_2\CausalState_5\CausalState_3\CausalState_7\CausalState_4\CausalState_1$] & (110010)$^*$    & 6H$_a$  \\ \hline
[$\CausalState_5\CausalState_2\CausalState_4\CausalState_0\CausalState_1\CausalState_3\CausalState_7\CausalState_6$] & (00011101)$^*$   & 8H$_a$  \\ \hline
[$\CausalState_2\CausalState_5\CausalState_3\CausalState_7\CausalState_6\CausalState_4\CausalState_0\CausalState_1$] & (11100010)$^*$   & 8H$_a$  \\ \hline
[$\CausalState_3\CausalState_6\CausalState_5$] & (011)$^*$ & 9R \\ \hline       
[$\CausalState_4\CausalState_1\CausalState_2$] & (100)$^*$ & 9R \\ \hline  
[$\CausalState_7\CausalState_6\CausalState_5\CausalState_3$]  & (0111)$^*$      & 12R \\ \hline 
[$\CausalState_0\CausalState_1\CausalState_2\CausalState_4$]  & (1000)$^*$      & 12R \\ \hline  
[$\CausalState_3\CausalState_6\CausalState_4\CausalState_0\CausalState_1$] & (00011)$^*$     & 15R  \\ \hline
[$\CausalState_4\CausalState_1\CausalState_3\CausalState_7\CausalState_6$] & (11100)$^*$     & 15R  \\ \hline
[$\CausalState_5\CausalState_2\CausalState_4\CausalState_0\CausalState_1\CausalState_3\CausalState_6$]  & (0001101)$^*$ & 21R$_a$ \\ \hline
[$\CausalState_2\CausalState_5\CausalState_3\CausalState_7\CausalState_6\CausalState_4\CausalState_1$]  & (1110010)$^*$ & 21R$_a$ \\ \hline
[$\CausalState_3\CausalState_6\CausalState_4\CausalState_0\CausalState_1\CausalState_2\CausalState_5$]  & (0001011)$^*$ & 21R$_b$ \\ \hline 
[$\CausalState_4\CausalState_1\CausalState_3\CausalState_7\CausalState_6\CausalState_5\CausalState_2$]  & (1110100)$^*$ & 21R$_b$ \\ \hline
\hline
\end{tabular}
\caption{
The $19$ CSCs on an $\Range = 3$ \eM. In the first column, we give the CSC
  and in the second we show the stacking sequence in the H{\"{a}}gg notation 
  implied by this CSC. If these CSCs are strongly represented on the \eM, then
  we can interpret them as crystal structure. The corresponding crystal
  structures are shown in the third column in the Ramsdell
  notation.~\cite{note3} Some CSCs come in pairs related by spin-inversion
  symmetry,~\cite{vc01} {\it i.e.} [$\CausalState_0$] and [$\CausalState_7$]
  are both 3C structure, differing only in chirality. In cases where the
  Ramsdell notation is identical for different structures, we have attached a
  subscript to distinguish them. We list the period-$8$ hexagonal structures
  with a subscript to differentiate them from the more common 8H structure
  (00001111). One must perform \eMSR\ at $\Range = 4$ to discover this latter
  8H structure.
  } 
\label{FMStructuresOneMr_3}
\end{center}
\end{table}

\subsubsection{Structural Interpretations}
\label{StructuralInterpretations}

We begin by noting that a purely crystalline structure is simply the repetition
of a sequence of MLs. This is realized on an \eM\ as a CSC with a $\CSCP = 1$.
That is, an \eM\ consisting of a single CSC repeats the same state sequence
endlessly, giving a periodic stacking sequence, which physically is some
crystal structure. It is therefore useful to catalog all of the possible CSCs
on an $\Range=3$ \eM, and this is done in Table~\ref{FMStructuresOneMr_3}.
There are $19$ CSCs on an $\Range =3$ \eM, and each can be thought of as a
crystal structure if that CSC is strongly represented. (These should be
verified by tracing them out on Fig.~\ref{fig:dB3.general}.)  

However, if a nearly perfect crystal has a few randomly inserted stacking
errors, these ``mistakes'' are physically an interruption of the regular
ordering of MLs. That is, some error occurs, but after a relatively short
distance the crystal returns to its regular stacking rule, thus restoring the
crystalline structure. This is realized on an \eM\ as a CSC with
$\CSCP (\rm{crystal}) \approx 1$ and another weakly represented CSC with
$\CSCP (\rm{fault}) \ll 1$. In this way, we interpret weakly represented
CSCs as faults. 

With this understanding in place, we note that an \eM\ can quite naturally 
accommodate more than one crystal structure. Each such CSC must have 
a $\CSCP (\rm{crystal}) \approx 1$, but they can be ``connected'' through a
weak CSC, ($\CSCP (\rm{fault}) \ll 1$). However, to interpret two CSCs as
crystalline structure, each must have $\CSCP (\rm{crystal}) \approx 1$, 
and therefore necessarily they \emph{do not} share a CS. (If they did, at
least one CSC could not have a $\CSCP (\rm{crystal}) \approx 1$.) Similarly, 
\eMs\ can accommodate more that one faulting structure. 

\subsection{Faulting Structures on \EMs}

As we did with crystal structures on an $\Range = 3$ \eM,
it is instructive to identify some of the more common faults on the most
general $\Range=3$ \eM. We will consider only 2H and 3C structures with
growth, deformation, and layer-displacement faults; but the extension to
other crystal and fault structures is straightforward. We will only give
the faulting structure on an \eM\ to first order in the faulting probability,
so that the basic graphical structure is clear. Thus, the connection with
the FM is valid only for weak faulting; which is consistent with the FM's
domain of applicability. The \eM, however, is valid for any degree of
disorder, it is only the connection to the FM that is limited to weak
faulting.

We also note that the faults on the \eM\ in this interpretation are such
that the occurrence of two adjacent faults is \emph{suppressed}. If the
probability of encountering a fault on a ML is (say) $p$, then the
probability of two adjacent faults is $p^2$. That is, in our
attempt to use the FM to interpret the structures captured by an \eM,
we ignore these higher-order terms. Thus, the issue of random versus
nonrandom faulting in polytypism is not addressed here. But, again, we
note that the \eM\ description quite naturally describes random, nonrandom,
and periodic faulting structures.~\cite{sk94} 

\subsubsection{Growth Faults}

\begin{figure}
\begin{center}
\resizebox{!}{1.75in}{\includegraphics{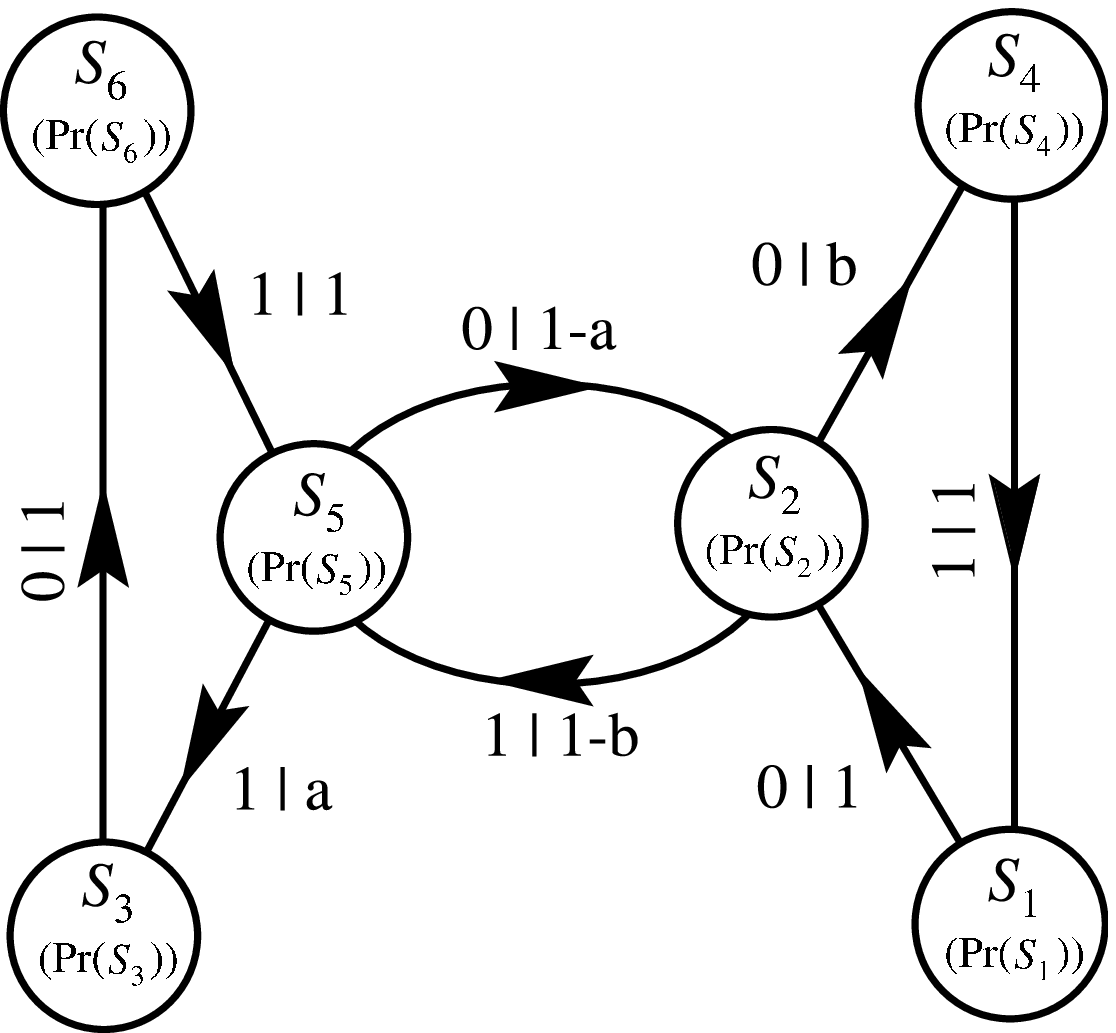}}
\end{center}
\caption{Growth faults for the 2H structure as they appear on a $\Range=3$ \eM\
  for small faulting probabilities $a$ and $b$. The
  [$\CausalState_2\CausalState_5$] CSC is the 2H structure and the CSCs
  [$\CausalState_5\CausalState_3\CausalState_6$]
  and [$\CausalState_2\CausalState_4\CausalState_1$] give the faulting. This
  interpretation is only valid for small faulting probabilities.}
\label{fig:dB3.growth.2H.S}
\end{figure}

Crystal growth often proceeds by a layer-addition process.
Suppose a ML is added that cannot be thought of as a continuation of
the previous crystal structure, but the MLs added subsequent to that ML
return to the original stacking rule. Such a ML inserted into the sequence
is called a \emph{growth fault}. For the 2H structure, the rule is that the
added ML has the same orientation as the next-to-last ML. For example,
imagine an unfaulted 2H crystal, consisting of $A$ and $B$ MLs, is
$...ABABAB...$. Then a growth fault in this structure is a $B$ ML 
followed by a $C$ ML. The remaining MLs continue to follow the 2H
stacking rule, giving an overall stacking sequence such as
\begin{eqnarray*}
   ...A\> B\> A\> B\> A\> B\> \underline{C}\> B\> C\> B\> C\> B... ~,
\label{fig:A}
\end{eqnarray*}
where underlining indicates the fault plane.

Notice that the original crystal is composed of alternating $A$ and $B$
MLs, while after the fault it becomes a sequence of alternating $B$
and $C$ MLs. In terms of the H\"{a}gg notation, a growth fault for the 2H
crystal corresponds to the insertion of a single $0$ or $1$ into the spin
sequence. For example, $...01010101...$ becomes $...0101\underline{1}0101...$
upon insertion of a $1$, where the underlining indicates the inserted spin.

This can be demonstrated on the \eM\ shown in Fig.~\ref{fig:dB3.growth.2H.S}
with small faulting probabilities $a$ and $b$. This \eM\ implies that the
CSC represented by [$\CausalState_2\CausalState_5$] is dominant, which is
simply the 2H structure. With small faulting probabilities $a$ or $b$,
a $0$ or a $1$, respectively, is randomly inserted into the 2H crystalline
structure.

\begin{figure}
\begin{center}
\resizebox{!}{1.75in}{\includegraphics{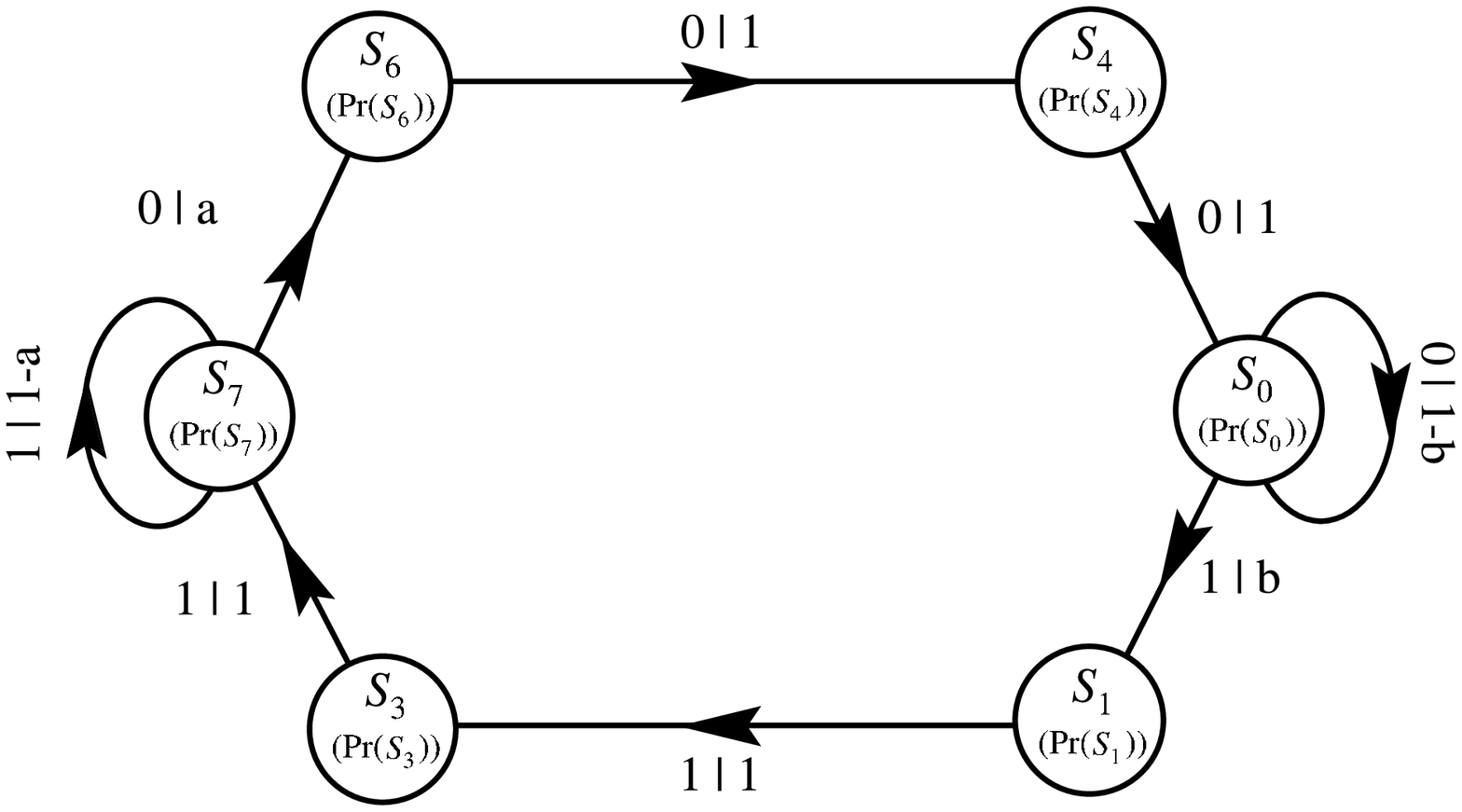}}
\end{center}
\caption{Growth faults for the 3C structure on a $\Range=3$ \eM\
  with small faulting probabilities $a$ and $b$. The CSCs [$\CausalState_7$] and
  [$\CausalState_0$] give the twinned 3C structure, while the paths
  $\CausalState_7\CausalState_6 \CausalState_4\CausalState_0$ and
  $\CausalState_0\CausalState_1\CausalState_3\CausalState_7$ give the faulting.
  Here we have an example of a faulting structure that induces a transition
  between two crystal structures.}
\label{fig:dB3.growth.3C.S}
\end{figure}

In the 3C structure, the stacking rule is that the added ML is
different from the previous two MLs. There are, of course, two
distinct, symmetry-related 3C structures; one being the $...ABCABC...$
and the other its spatial inversion $...CBACBA...$. The spin sequences
for these are $...1111...$ and $...0000...$, respectively. A growth fault
for this crystal gives a stacking sequence such as
\begin{eqnarray*}
   ...A\> B\> C\> A\> B\> \underline{C}\> B\> A\> C\> B\> A... ~,
\label{fig:B}
\end{eqnarray*}
where underlining again indicates the fault plane. It is conventional to take
the indicated ML as the fault plane since it is the only ML in the cubic
stacking sequence that is hexagonally related to its neighbors. In terms of
H\"{a}gg notation, the sequence is $...11111|00000...$, where the vertical
line indicates the fault plane. The effect of a growth fault in a 3C
structure is then to switch from 3C structure of one chirality to another
or to flip all spins after the fault plane.  This fault is also known as a
\emph{twin fault} of the 3C structure, because it produces a crystal containing
both kinds of 3C sequences. This growth fault is demonstrated in
Fig.~\ref{fig:dB3.growth.3C.S} with small faulting probabilities $a$ and $b$.
The two CSCs [$\CausalState_7$] and [$\CausalState_0$] correspond to the two
twinned 3C structures, with the transition sequences connecting them having a
small total probability.

\subsubsection{Deformation Faults}

\begin{figure}
\begin{center}
\resizebox{!}{1.75in}{\includegraphics{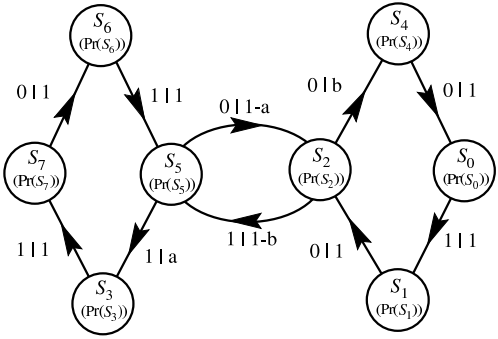}}
\end{center}
\caption{Deformation faulting for the 2H structure on a $\Range=3$
  \eM\ for small fault probabilities $a$ and $b$. A deformation fault
  is represented by a single spin flip.}
\label{fig:dB3.deformation.2H.S}
\end{figure}

Other faults can occur after a crystal structure has been formed. Caused
by external stresses or inhomogeneous temperature distributions within
the crystal, \emph{deformation faults} are the result of one plane in the
crystal slipping past another in a direction transverse to the stacking.
An example of deformation faulting in the 2H structure is the following:
\begin{eqnarray*}
   ...A\> B\> A\> B\> |\> C\> A\> C\> A\> C\> A... ~.
\label{fig:C}
\end{eqnarray*}
The vertical bar indicates the plane across which the slip
occurred. In the H\"{a}gg notation a deformation fault in the 2H structure
is realized by flipping a spin. In this example, the unfaulted sequence
...10101010... transforms to ...101\underline{1}1010..., where again
the underlined spin demarcates the one flipped. The \eM\ representation
of this fault is shown in Fig.~\ref{fig:dB3.deformation.2H.S}.

\begin{figure}
\begin{center}
\resizebox{!}{1.75in}{\includegraphics{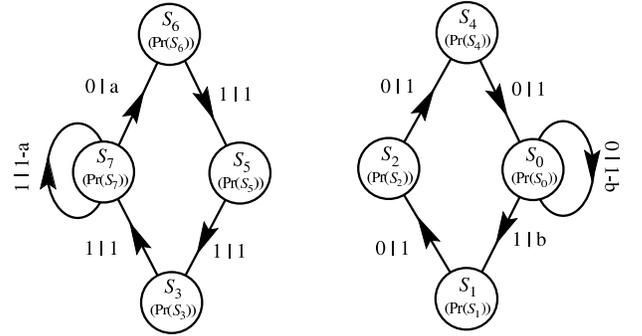}}
\end{center}
\caption{The two possible \eMs\ for deformation faulting in the 3C structure
  with small faulting probabilities $a$ and $b$. 
  There are two \eMs, one for faulting from each of the 3C structures,
  [$\CausalState_7$] and [$\CausalState_0$]. They are disconnected, and
  hence faulting in the 3C structure of one chirality cannot cause
  the crystal to switch to another chirality. Hence this faulting
  mechanism does not cause twinning.} 
\label{fig:dB3.deformation.3C.S}
\end{figure}

In the 3C structure, deformation faults appear much the same. An example
of a deformation fault in a 3C structure is
\begin{eqnarray*}
   ...A\> B\> C\> A\> B\> C\> |\> B\> C\> A\> B\> C\> A...
\label{fig:D}
\end{eqnarray*}
The vertical bar again indicates the slip plane. Expressed in spin notation,
the unfaulted 3C crystal $...11111111...$ becomes $...1111\underline{0}111...$,
a single spin flip. The two corresponding \eMs\ are shown in
Fig.~\ref{fig:dB3.deformation.3C.S}.

\subsubsection{Layer-Displacement Faults}

\begin{figure}
\begin{center}
\resizebox{!}{1.75in}{\includegraphics{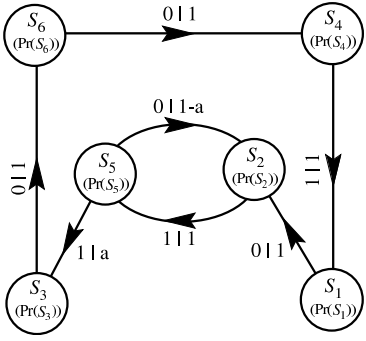}}
\end{center}
\caption{Layer-displacement faults for the 2H structure on a $\Range=3$
  \eM\ with small faulting probability $a$. Here, for the sake of clarity,
  we show only faulting initiating from the $\CausalState_5$ CS, although
  a similar faulting structure can begin from the $\CausalState_2$ CS.}
\label{fig:dB3.layer.displacement.2H.S}
\end{figure}

\emph{Layer-displacement faults} are characterized by a shifting 
of one or two MLs in the crystal, while leaving the remainder of the crystal
undisturbed. As such, these faults do not disrupt the long-range order
present in a structure. They are thought to be introduced at high
temperatures by diffusion of the atoms through the crystal.~\cite{sk94}
In the 2H structure, an example of a layer-displacement fault is:
\begin{eqnarray*}
   ...A\> B\> A\> B\> \underline{C}\>  B\> A\> B\> A... ~,
\label{fig:E}
\end{eqnarray*}
where the underlined ML is the faulted layer. Written as spins,
$...10101010...$ becomes $...101\underline{10}010...$, the underlined
spins indicating those that have flipped. A layer-displacement fault
in the 2H structure is shown in Fig.~\ref{fig:dB3.layer.displacement.2H.S}.

\begin{figure}
\begin{center}
\resizebox{!}{1.75in}{\includegraphics{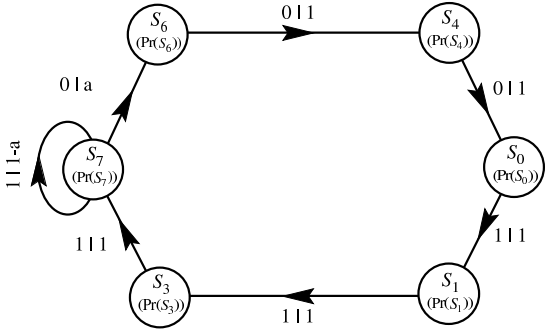}}
\end{center}
\caption{Layer-displacement faults for the 3C structure on a $\Range=3$
  \eM\ with small faulting probability $a$. Again, we show only
  faulting from the $\CausalState_7$ CS, corresponding to the $...1111...$
  spin sequence. A similar fault structure could be drawn for faulting
  from the $\CausalState_0$ CS. A layer-displacement fault for the 3C
  structure is achieved by three consecutive spin flips. The difference
  between the fault structure here and that of a growth fault for the 3C
  structure is that growth faults produce twinning, whereas here the fault
  returns to the original crystal structure.}
\label{fig:dB3.layer.displacement.3C.S}
\end{figure}

Layer-displacement faults in 3C structures are more difficult to realize,
since each ML is sandwiched between two unlike MLs and changing
its orientation violates the stacking constraints. It is therefore
necessary for two adjacent MLs to shift. Consequently, one expects
that they are rarer. An example of layer-displacement in the 3C
structure is the following:
\begin{eqnarray*}
   \ldots
   A\> B\> C\> A\> B\> C\> \underline{B\> A}\> C\> A\> B\> C\> A B\>
   \ldots ~,
\label{fig:F}
\end{eqnarray*}
where the underlined layers are faulted. The spin sequence changes from
$\ldots 1111111 \ldots$ to one where three consecutive spins have been
flipped to $0$: $\ldots ...1100011 \ldots$. A layer-displacement fault
on the 3C structure is shown in Fig.~\ref{fig:dB3.layer.displacement.3C.S}.

These common faulting structures for the 2H and 3C crystals are given in Table~\ref{FMfaultsOneMr_3} 
along with the CSCs associated with them. 

\begin{table}
\begin{center}
\begin{tabular}{l|l|l}
\hline
\hline
2H  & Growth fault  & $\CausalState_5\CausalState_3\CausalState_6$  \\ \hline   
    &               & $\CausalState_2\CausalState_4\CausalState_1$  \\ \hline
    & Deformation fault & $\CausalState_5\CausalState_3\CausalState_7\CausalState_6$  \\ \hline 
    &                   & $\CausalState_2\CausalState_4\CausalState_0\CausalState_1$  \\ \hline
    & Layer-displacement fault & $\CausalState_5\CausalState_3\CausalState_6\CausalState_4\CausalState_1\CausalState_2$  \\ \hline
    &                          & $\CausalState_2\CausalState_4\CausalState_1\CausalState_3\CausalState_6\CausalState_5$  \\ \hline
\hline
3C  & Growth fault  & $\CausalState_7\CausalState_6\CausalState_4\CausalState_0$  \\ \hline
    &               & $\CausalState_0\CausalState_1\CausalState_3\CausalState_7$  \\ \hline
    & Deformation fault & $\CausalState_7\CausalState_6\CausalState_5\CausalState_3$  \\ \hline
    &                   & $\CausalState_0\CausalState_1\CausalState_2\CausalState_4$  \\ \hline
    & Layer-displacement fault & $\CausalState_7\CausalState_6\CausalState_4\CausalState_0\CausalState_1\CausalState_3$  \\ \hline
    &                          & $\CausalState_0\CausalState_1\CausalState_3\CausalState_7\CausalState_6\CausalState_4$  \\ \hline
\hline
\end{tabular}
\caption{The more common fault structures for the 2H and 3C structures on an
  $\Range = 3$ \eM. We make the following interpretation: If there is one
  parent structure (2H or 3C) that is strongly represented and a single
  additional CSC is associated with it as shown above, then we say that that
  crystal has the structure of the parent crystal with a certain amount 
  of the given faulting. We should be clear here, however, not to confuse
  structure with mechanism. In this interpretation, the \eM\ gives the
  structure that a crystal would have {\em if} it experienced a small amount
  of the faulting given. Structure does not necessarily imply mechanism. 
  }
\label{FMfaultsOneMr_3}
\end{center}
\end{table}

\subsection{\EM\ Decomposition and General Interpretation}

The previous discussion has emphasized the important r{\^{o}}le that CSCs 
play in reflecting stacking structures---crystalline and fault.  
We have found that CSCs directly correspond to crystal and fault 
structures. The question then arises, can any arbitrary $\Range = 3$ \eM\
be decomposed into crystal and fault structures? We first note that the
\emph{only} difference between fault and crystal structure is the magnitude
of the $\CSCP$ associated with each CSC. It seems reasonable, then, to break
an \eM\ into a sum of CSCs. So we formally write,
\begin{eqnarray}
    \mathcal{E} \sim \sum_i \nu_i (CSC_i) ~,
\label{eq:e_machine}
\end{eqnarray}
where $\mathcal{E}$ is the \eM, $\nu_i$ is the
fraction of the \eM\ that can be attributed to the
$i^{\mathrm{th}}$ CSC, and $CSC_i$ is the $i^{\mathrm{th}}$ CSC. The
most general binary $\Range = 3$ \eM\ can be specified by
eight variables. It is known, however, that there are $19$ CSCs on such
an \eM.~\cite{t90} So, unless there is a fortuitous vanishing of
either CSs or \eM\ transitions, or the imposition of additional constraints,
the decomposition in Eq.~(\ref{eq:e_machine}) is \emph{not} unique
and therefore of questionable use. We note this situation is not expected
to improve with larger $\Range$. The number of parameters on the
$\Range^{\mathrm{th}}$ \eM\ grows exponentially in $\Range$,
while the number of CSCs appears to increase as the exponential of
an exponential in $\Range$.~\cite{t90} Thus, in general, it is not possible
to decompose an \eM\ into CSCs uniquely.

The main purpose of such a decomposition is to provide intuition into the
structure present and possibly insight into the
physical mechanisms that may have led to a particular structure. In 
this limited capacity Eq.~(\ref{eq:e_machine}) may be helpful. We
stress that Eq.~(\ref{eq:e_machine}) has no other use than this and
certainly cannot be used to calculate physical quantities. Only the entire
\eM\ is suitable for such calculations.

How, then do we intuitively understand structure on an \eM? In one
picture---the weak faulting limit---we  view the \eM\ as a collection of CSCs.
The decomposition given by Eq.~(\ref{eq:e_machine}), while not unique, may be
sensible. In this case, we have the same interpretation as the FM. However, 
the \eM\ has a broader range of applicability. It can, for instance,
accommodate more than one crystal structure and detail how the stacking
alternates between the two. The FM, to our knowledge, admits no such
multicrystalline structures. The \eM\ provides a more detailed account of
multiple faulting structures. The FM, too, can reflect more than one kind of
faulting structure, but the description is a stochastic one. The \eM\ has no
difficulty in reflecting any stacking structures, even closely spaced faulting
structures. Thus, we retain the structural interpretations of
\S \ref{StructuralInterpretations} for the case of weakly faulted structures.  

In other circumstances, when a sensible decomposition of the \eM\ into crystal
and faulting structures is not possible, it still gives insight into important
stacking sequences and their spatial relations. Although it is no longer as
advantageous to view the \eM\ as a collection of CSCs, we note that stacking
sequence probabilities are readily observed on the \eM\ either through direct
calculation---the \eM\ specifies sequence frequencies of any length---or, more
simply for shorter sequences, by asymptotic CS probabilities. The likelihood
of seeing two sequences in close proximity can be found by tracing the
appropriate path through the \eM. Since the \eM\ is valid for \emph{any}
degree of disorder, we can find the relative importance of stacking sequences
for even heavily faulted crystals or for crystals in which no regular stacking
structures exist. The architecture of the \eM---{\it i.e.}, the number,
arrangement, and connections between the CSs, then provides an intuitive
interpretation for the complexity and organization of the structure. One sees
how the various stacking structures are related and how one blends into 
another upon scanning the crystal.  

Thus, in addition to providing a formidable calculational tool, the
\eM\ provides a new way of viewing structure in layered materials; it is not
tethered to the assumption of a parent crystal permeated with weak faults.
It gives a generalized way to view and compare the structure of different
crystals, even when they have different---or no---parent structures. This
should prove especially helpful in understanding solid-state transformations
in layered materials, as we intend to demonstrate in follow-on work.  
  
Finally, we note that these are interpretations of convenience, not necessity. 
The \eM\ is a unique description of the stacking process, and thus any
quantities that depend directly on a statistical description of the stacking
are amenable to calculation.  

For those instances where a sensible decomposition of the \eM\ is
possible---{\it i.e.}, the weak faulting limits---we employ
Eq.~(\ref{eq:e_machine}) for the limited purpose of providing an intuitive
understanding of the disordered structure  We will call $\nu_i$ either
the \emph{fraction} of crystal structure or, for weak CSCs, the
\emph{fault density}. This is, of course, different from the
\emph{fault probability} generally used in the literature. The fault
probability is the frequency, upon scanning the stacking sequence, that one
finds a particular fault in the sequence.

\section{Example Processes}
\label{ExampleProcesses}

We consider four prototype processes to demonstrate \eMSR. In Example A, we
reconstruct an \eM\ for a known $\Range = 3$ process and show that the
technique works in this case and, indeed, for any process that has
$\MLength \leq 3$. In Example B, we consider a process that cannot be
represented on a $\Range = 3$ \eM. We reconstruct an
\eM\ for a process that requires a $\MLength = 4$ memory length and 
find that a $\Range = 3$ \eM\ gives a reasonable approximation.
In Example C, we treat a process with $\MLength = 1$ to demonstrate  
that \eMSR\ does terminate at the minimum $\Range$. We also show 
that had we not terminated reconstruction at $\Range = 1$, the 
equivalence relation, Eq. (\ref{eq:cm}) would require the merging of
equivalent histories that would effectively find the $\Range = 1$ \eM.  
Finally, in Example D, we reconstruct the $\Range = 3$ \eM\ for the
even process---a finite-state process that has a distinctive kind
of infinite memory.~\cite{cf01,c92} Again, we find that the $\Range = 3$
\eM\ gives a reasonable approximation. 

In Examples A, B, and D, we solve the
spectral equations at $\Range = 3$ with the memory-length reduction 
approximation via a Monte Carlo technique.~\cite{v01} These equations are
given in Appendix \ref{SpectralEquations}. To find the predicted CFs
for each \eM, we take a sample spin sequence generated by the \eM\ of
length 400 000 and find the CFs by directly scanning the resulting stacking
sequence. The diffraction spectra are calculated from Eq. (\ref{eq:diff1})
using a sample of 10 000 MLs. Since these are theoretical spectra, and have
no error, we are not able to set an acceptable threshold error in advance.
Instead, we have chosen examples, except for Example C, that require the
$\Range = 3$ solutions and we solve these examples at $\Range = 3$. In the
event that a CS is assigned an asymptotic state probability of less than
$0.01$, we take that CS to be nonexistent.

We also calculate the information-theoretic quantities described in
\S \ref{MeasuresStructureIntrinsicComputation} for each example and the
reconstructed \eM\ and display the results in Table~\ref{tab:results}.
Analyzing these examples not only demonstrates the feasibility and
accuracy of spectral \eM\ reconstruction, but they also serve to
illustrate how \eMs\ capture structure and disorder. In the companion
sequel,~\cite{vcc02c} we apply the same procedures to the
analysis of experimental diffraction spectra from ZnS, focusing on
the novel physical and material properties that can be discovered
with this technique.

\subsection{Example A}

\begin{figure}
\begin{center}
\resizebox{!}{1.75in}{\includegraphics{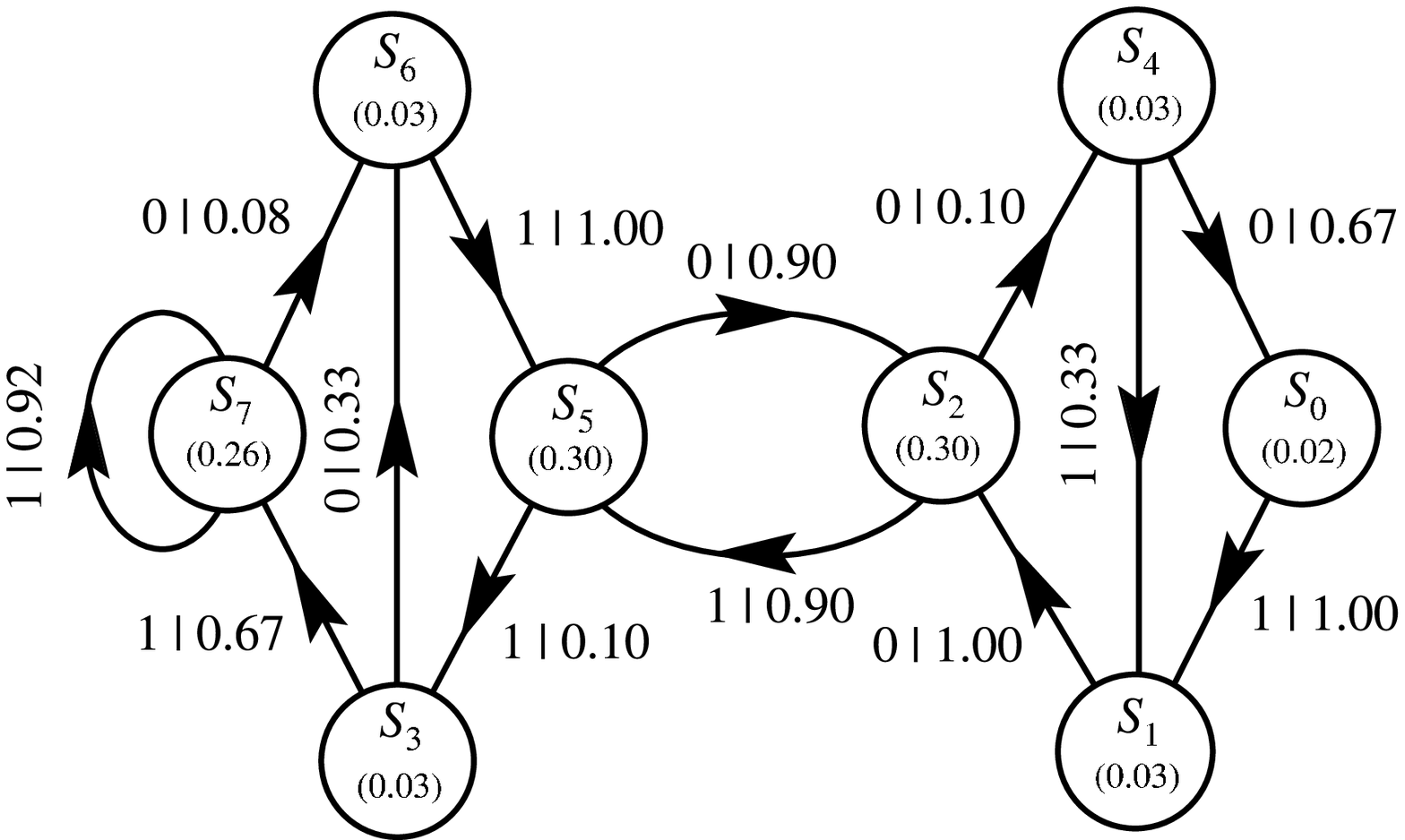}}
\end{center}
\caption{The $\Range = 3$ \eM\ for the Example A process. The asymptotic
  probabilities are given for each CS. The large CSC probabilities
  for the [$\CausalState_7$] CSC ($\CSCP([\CausalState_7]) = 0.92$) and the 
  [$\CausalState_2\CausalState_5$] CSC ($\CSCP[\CausalState_2\CausalState_5]) = 0.81$) 
  suggest that one think of these cycles as crystal structure and everything
  else as faulting.
  }
\label{fig:dB3.example.A}
\end{figure}

\begin{figure}
\begin{center}
\resizebox{!}{5cm}{\includegraphics{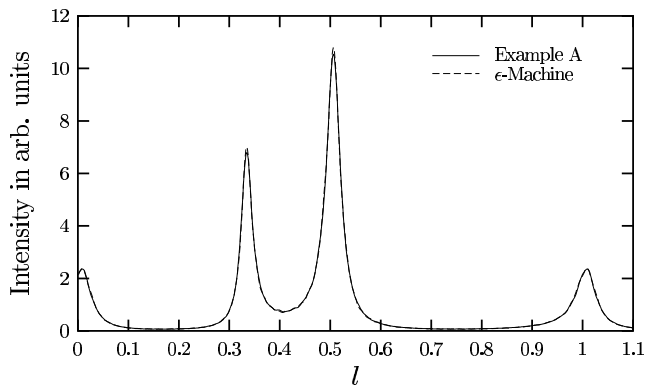}}
\end{center}
\caption{A comparison between the diffraction spectra $\mathsf{I}(l)$
  generated by Example A and by the $\Range = 3$ spectrally reconstructed
  \eM. The differences between the diffraction spectra for Example A and the
  $\Range = 3$ reconstructed \eM\ are too small to be seen. We calculate
  ${\cal R} = 2 \%$, but this is largely due to numerical error. (See text.) 
  The peak at $l \approx 1/3$ corresponds to the 3C structure and the two peaks
  at $l \approx 1/2$ and $l \approx 1$ to the 2H structure.
  }
\label{fig:dp.example.a}
\end{figure}

We begin with the sample process given in Fig.~\ref{fig:dB3.example.A}.
This process can approximately be decomposed into FM structural components
using Eq.~(\ref{eq:e_machine}) in the following way:
\begin{center} \(
\begin{array}{lr}
   \textrm{2H}                            &    $ 54\%         $   \\
   \textrm{3C$^+$}                        &    $ 24\%         $   \\
   \textrm{Deformation fault}             &    $ 16\%         $   \\
   \textrm{Growth fault}                  &    $  6\%         $
\end{array}
\) \end{center}
where the ``+'' on 3C indicates that only the positive chirality
($...1111...$) structure is present. The faulting is given with reference
to the 2H crystal.

The diffraction spectrum from this process is shown in
Fig.~\ref{fig:dp.example.a}. The experienced crystallographer has little
difficulty guessing the underlying crystal structure: the peaks at
$l \approx 1/2$ and at $l \approx 1$ suggest the 2H structure;
while the peak at $l \approx 1/3$ is characteristic of the 3C structure.

The mechanism responsible for the faulting is less clear, however. It is
known that various kinds of fault
produce different effects on the Bragg peaks.~\cite{sk94} For instance,
both growth and deformation faults broaden the peaks in the diffraction
spectrum of the 2H structure, the difference being that growth faults
broaden the integer-$l$ peaks three times more than the half-integer-$l$
peaks, while broadening due to deformation faulting is about equal. The
FWHM for the peaks are $0.028$, $0.034$, and $0.049$ for $l \approx 0.33$,
$0.5$, and $1$, respectively. This gives then a ratio of about $1.4$
for the integer-$l$ to half-integer-$l$ broadening, suggesting (perhaps)
that deformation faulting is prominent. One expects there to be no shift
in the position of the peaks for either growth or deformation faulting;
which is clearly not the case here. In fact, the two peaks associated with
the 2H structure at $l \approx 0.5$ and 1 are shifted by
$\Delta l \approx 0.006$ and 0.009, respectively.
This analysis is, of course, only justified for one parent crystal in the
overall structure, nonetheless if we neglect the peak shifts, the simple  
intuitive analysis appears to give good qualitative results here.

With the 3C peak, both deformation and growth faults produce a broadening, the
difference being that the broadening is asymmetrical for the growth faults.
One also expects there to be some peak shifting for the deformation faulting.
There is a slight shift ($\Delta l \approx 0.002$) for the $l \approx 1/3$
peak and the broadening seems (arguably) symmetric, so one is tempted to guess
that deformation faulting is important here.  Indeed, the CSC
[$\CausalState_7\CausalState_6\CausalState_5\CausalState_3$] is consistent
with deformation faulting in the 3C crystal. Heuristic arguments, while not
justified here, seem to give qualitative agreement with the known structure.

\begin{figure}
\begin{center}
\resizebox{!}{5cm}{\includegraphics{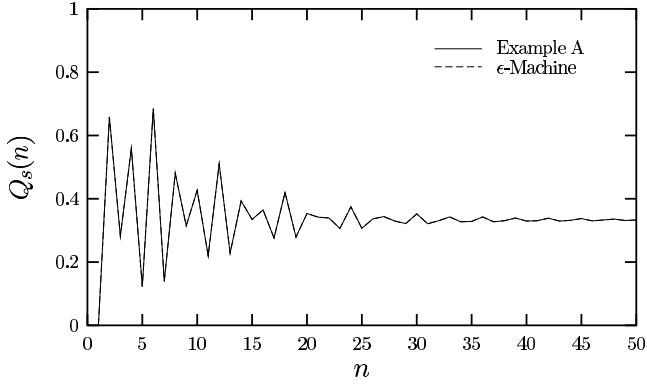}}
\end{center}
\caption{A comparison of the CFs $Q_s(n)$ between the Example A process
  and the $\Range = 3$ reconstructed \eM.  As with the diffraction
  spectra, the differences are too small to be seen on the graph. As an aid
  to the eye, here and in other graphs showing CFs, we connect the the values
  of adjacent CFs with straight lines. The CFs, of course, are defined only
  for integer values of $n$.
  }
\label{fig:qs.example.a}
\end{figure}

The \eM\ description does better. We follow the spectral reconstruction
procedure given in \S \ref{SpectralEMReconstruction}. We Fourier analyze
the spectrum over the interval $ 0 \le l \le 1$. The figures-of-merit are
equal to their theoretical values within numerical error. The reconstructed
\eM\ is equivalent to the original one, with CS probabilities and transition
probabilities typically within 0.1\% of their original values, except for the
transition $1 | 0.33$ from $\CausalState_4$, which was 1\% too small. Not
surprisingly, the process shown in Fig.~\ref{fig:dB3.example.A} \emph{is}
the reconstructed \eM\ and so we do not repeat the figure.

The two-layer CFs $Q_s(n)$ versus $n$ from the process and from the
reconstructed \eM\ are shown in Fig.~\ref{fig:qs.example.a}. The differences
are too small to be seen on the graph. We calculate the profile $\cal R$-factor
to compare the ``experimental'' spectrum (Example A) to the ``theoretical''
spectrum (\eM) and find a value of ${\cal R} \approx 2 \%$. If we generate
several spectra from the same process, we find profile $\cal R$-factors of
similar magnitude. This error is then must be due to sampling. It stems from
the finite spin sequence length we use to calculate the CFs and our method
for setting them equal to their asymptotic value. (See \S
\ref{EstimateStackingSequenceDistribution}.) This can be improved  by taking
longer sample sequence lengths and refining the procedure for setting the
CFs to their asymptotic value. Since typical profile $\cal R$-factors
comparing theory and experiment are much larger than this, at present, this
does not seem problematic. A comparison of the two spectra is shown in Fig.
(\ref{fig:dp.example.a}). This kind of agreement is typical of spectral
\eM\ reconstruction from any process that can be represented as a
$\Range = 3$ \eM.~\cite{v01}

We find by direct calculation from the \eM\ that both Example A and the
reconstructed process have a configurational entropy of $\hmu \approx 0.44$
bits/spin, a statistical complexity of $\Cmu \approx 2.27$ bits, and an
excess entropy of $\EE \approx 0.95$ bits.

Since the original process was representable as an $r = 3$ \eM, this first
example is largely a consistency check on \eMSR. In the next example, we treat
an $\Range > 3$ process not representable by the $\Range = 3$ \eMs\ that we
reconstruct.

\subsection{Example B}

\begin{figure}
\begin{center}
\resizebox{!}{1.75in}{\includegraphics{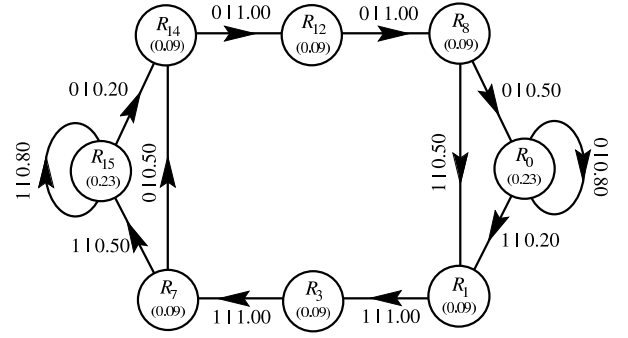}}
\end{center}
\caption{The process for Example B. Since it has a memory of $\MLength = 4$,
  we label the states with the last four spins observed: i.e.,
  $\mathcal{R}_{12}$ means that $1100$ were the last four spins. The
  CSCs [$\mathcal{R}_{15}$] and [$\mathcal{R}_0$] give rise to 3C structure
  and the CSC
  [$\mathcal{R}_1\mathcal{R}_3\mathcal{R}_7\mathcal{R}_{14}\mathcal{R}_{12}\mathcal{R}_8$]
  generates 6H structure.}
\label{fig:dB4.example.B}
\end{figure}

Upon annealing, a solid-state transformation in ZnS from the 2H structure
to either the 3C or 6H structures is possible, sometimes both occurring
in different parts of the same crystal.~\cite{sk94} However, two crystal
structures represented with an \eM\ cannot share a CS, as discussed in
\S \ref{StructuralInterpretations}. On a $\Range = 3$ \eM, for example,
both the CSCs associated with the 3C and the 6H structures share the CSs
$\CausalState_7$ and $\CausalState_0$, so a crystal containing both structures
cannot be properly modeled at $\Range = 3$. In fact, it is necessary to use
an $\Range = 4$ \eM\ to encompass both structures. So, to see how well spectral
reconstruction works at $\Range = 3$ for an $\Range = 4$ process, we consider
the process shown in Fig.~\ref{fig:dB4.example.B}. The CSC
[$\mathcal{R}_1\mathcal{R}_3\mathcal{R}_7\mathcal{R}_{14}\mathcal{R}_{12}\mathcal{R}_8$]
would give rise to 6H structure if it were a strong CSC, but we find that
$P_{6H} = 0.25$. We say then that this is mild 6H structure.   
The CSCs [$\mathcal{R}_0$] and
[$\mathcal{R}_{15}$] give the twinned 3C structures.

\begin{figure}
\begin{center}
\resizebox{!}{1.75in}{\includegraphics{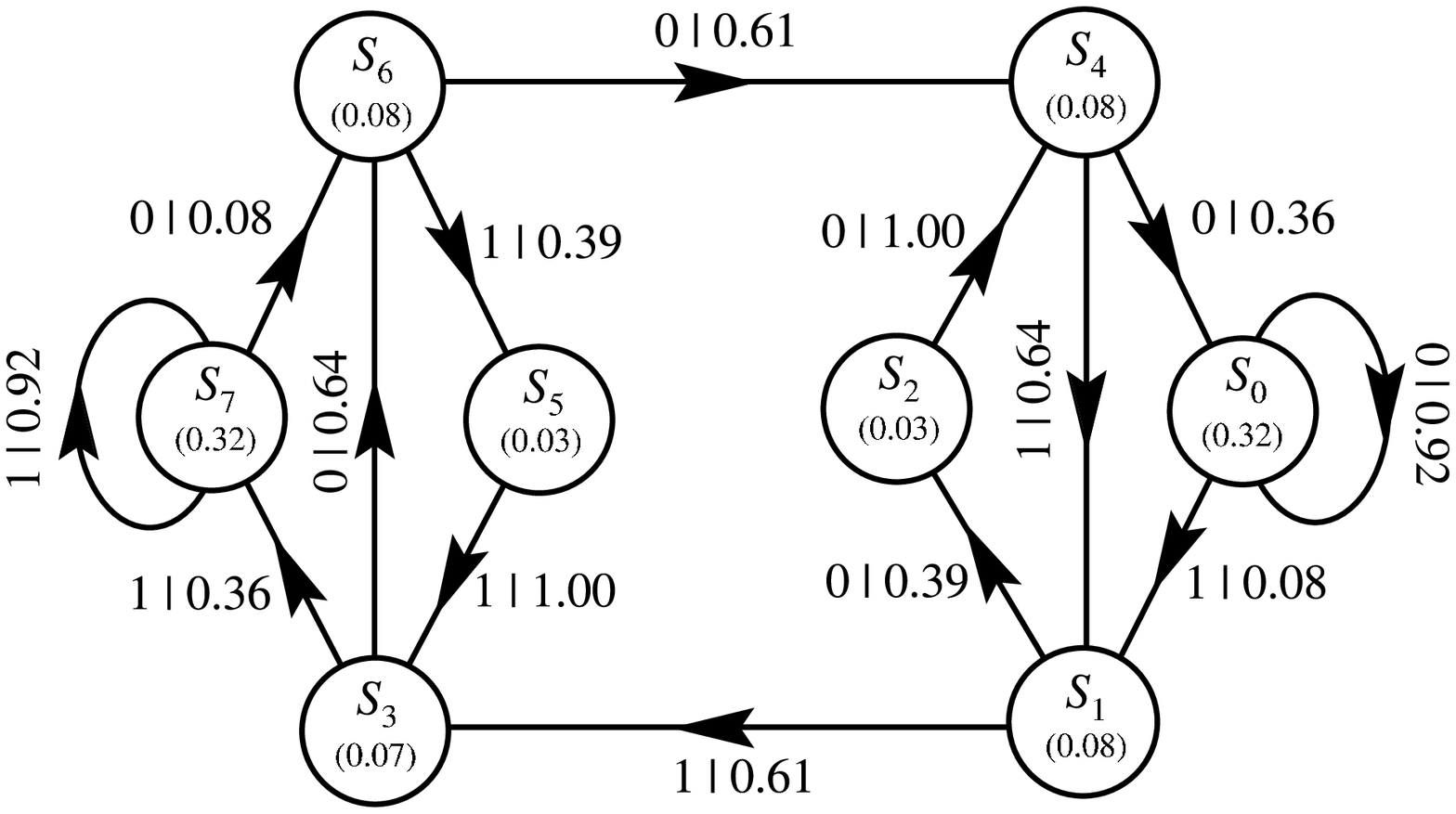}}
\end{center}
\caption{The \eM\ reconstructed at $\Range = 3$ for Example B.}
\label{fig:dB3.example.B}
\end{figure}

\begin{figure}
\begin{center}
\resizebox{!}{5cm}{\includegraphics{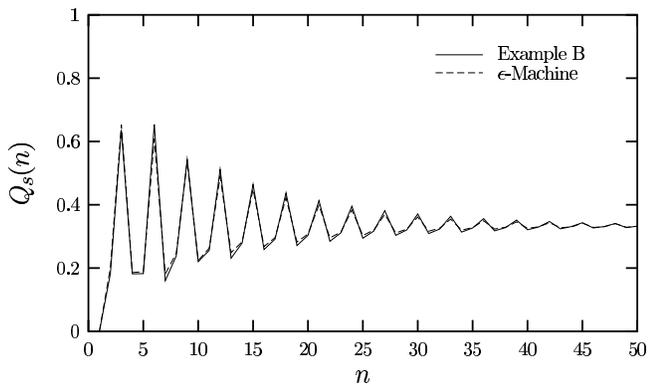}}
\end{center}
\caption{A comparison of the CFs $Q_s(n)$ generated by the $\Range = 3$
  reconstructed \eM\ (dashed line) and generated by Example B (solid line).
  The agreement is excellent.
  }
\label{fig:qs.example.B}
\end{figure}

The figures of merit were all equal to their theoretical values within
numerical error. Employing spectral reconstruction, we find the $\Range = 3$
\eM\ shown in Fig.~\ref{fig:dB3.example.B}. All CSs are present and all
transitions, save those that connect the $\CausalState_2$ and $\CausalState_5$
CSs, are present. A comparison of the CFs for the original process and the
reconstructed \eM\ is given in Fig.~\ref{fig:qs.example.B}. The agreement is
remarkably good. It seems that the $\Range = 3$ \eM\ picks up most of the
structure in the original process.

\begin{figure}
\begin{center}
\resizebox{!}{5cm}{\includegraphics{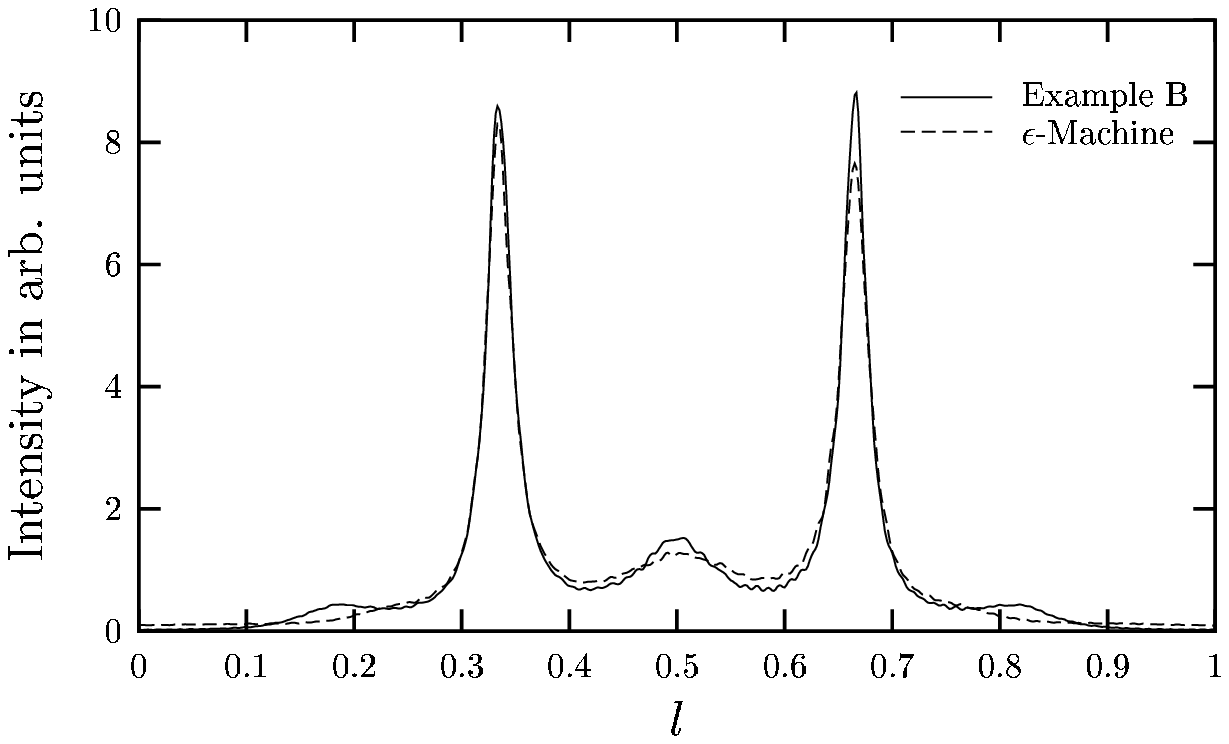}}
\end{center}
\caption{A comparison of the diffraction spectra $\mathsf{I}(l)$ between
  $\Range = 3$ reconstructed \eM\ and the process of Example B. The agreement
  is surprisingly good. The small peaks at $l \approx 1/6$ and $l \approx 5/6$
  correspond to the 6H structure. The $\Range = 3$ \eM\ has difficulty in
  reproducing these because the 6H and the 3C structure both share the
  $\CausalState_7$ and $\CausalState_0$ CSs and so require an \eM\ reconstructed
  at $\Range = 4$ to properly disambiguate them.
  }
\label{fig:dp.example.B}
\end{figure}

There is similar, though not as
good, agreement in the diffraction spectra, as Fig.~\ref{fig:dp.example.B}
shows. The most notable discrepancies are in the small rises at
$l \approx 1/6$ and $l \approx 5/6$.
We calculate a profile ${\cal R}$-factor of ${\cal R} = 12\%$ between the
diffraction spectra for Example B and the reconstructed \eM.
The $\Range = 3$ \eM\ has difficulty
reproducing the 6H structure in the presence of 3C structure, as
expected.

Given the good agreement between the correlation functions and the spectra
generated by Example B and the $\Range = 3$ \eM, we are led to ask what the
differences between the two are. In Table~\ref{tab:word_probs_3B} we give
the frequencies of the eight length-3 sequences generated by each process.
The agreement is excellent. They both give the same probabilities for the most
common length-$3$ sequences, $111$ and $000$. Example B does forbid two
length-$3$ sequences, $101$ and $010$, which the reconstructed $\Range = 3$
\eM\ allows with a small probability (0.03). At the level of length-$3$
sequences, the \eM\ is capturing most of the structure in the stacking
sequence. 

A similar analysis allows us to compare the probabilities of the $16$
length-$4$ sequences generated by each; the results are given in
Table~\ref{tab:word_probs_4B}. There are more striking differences here. The
frequencies of the two most common length-4 sequences in Example B,
$\Prob(1111) = \Prob(0000) = 0.23$, are overestimated by the $\Range = 3$
\eM, which assigns them a probability of $0.29$ each. Similarly, sequences
forbidden by Example B---$1101$, $1011$, $1010$, $1001$, $0110$, $0101$, $0100$,
$0010$---are not necessarily forbidden by the $\Range = 3$ \eM. In fact, the
$\Range = 3$ \eM\ forbids only two of them, $0101$ and $1010$. This implies
that $\Range = 3$ \eM\ can find spurious sequences that are not in the original
stacking sequence. This is to be expected. But the $\Range = 3$ \eM\ {\em does}
detect important features of the original process. It finds that this is a
twinned 3C structure. It also finds that 2H structure plays no role in the
stacking process. (We see this by the absence of transitions between the
$\CausalState_2$ and $\CausalState_5$ CSs in Fig.~\ref{fig:dB3.example.B}.)   

One can also attempt to decompose the $\Range = 3$ \eM\ into a sum of CSCs and
interpret this as crystal and fault structure. However, as is typically the
case, there is no unique decomposition and so therefore such an exercise is of
questionable validity. With the exception of the sequences $1111$ and $0000$,
the other twelve nonvanishing sequences all appear with a small, but rather
constant probability of $0.03$ or $0.05$. One possible interpretation is to
say that the CSCs [$\CausalState_0$] and [$\CausalState_7$] contribute to 3C
structure with a weight of $0.58$. We could further interpret the
[$\CausalState_7\CausalState_6\CausalState_5\CausalState_3$] and
[$\CausalState_0\CausalState_1\CausalState_2\CausalState_4$] CSCs as
deformation faulting of the 3C structure giving a combined weight of $0.24$.
And finally, we could associate the CSC
[$\CausalState_1\CausalState_3\CausalState_6\CausalState_4$] with 4H structure.
This last interpretation of the CSC
[$\CausalState_1\CausalState_3\CausalState_6\CausalState_4$] with any crystal
structure is troublesome as the $P_{CSC} \ll 1$. Another possible decomposition
would be to again assign the CSCs [$\CausalState_7$] and [$\CausalState_7$]
to the 3C structure with a weight of 0.58, to interpret the paths 
$\CausalState_7$$\CausalState_6$$\CausalState_4$$\CausalState_0$ and
$\CausalState_0$$\CausalState_1$$\CausalState_3$$\CausalState_7$ as
as twin faulting with a probability weight of $0.18$, treat the CSC
[$\CausalState_1\CausalState_3\CausalState_6\CausalState_4$] as 4H structure,
and finally to interpret the two CSCs
[$\CausalState_1\CausalState_2\CausalState_4$] and
[$\CausalState_3\CausalState_6\CausalState_5$] as 9R structures. These two
descriptions are clearly rather different and, arguably, have no use in any
account, other than serving to illustrate the ambiguity of FM-like
structural interpretations. 

In addition to the nonuniqueness difficulties, by simply listing the
probability density of the various crystals and fault structures, 
we say nothing about how one crystal converts into another as one 
scans the stacking sequence. This exercise demonstrates the impoverished
view of crystal structure inherent in the FM. In short, the stacking
sequence implied by the \eM\ in Fig.~\ref{fig:dB3.example.B} comes
from a physical structure that is not describable in terms of the FM. 

\begin{table}
\begin{center}
\begin{tabular}{lcc||lcc}
\hline \hline
Sequence   & Example B & \eMSR\ & Sequence   & Example B & \eMSR\ \\
\hline
111  &  0.32   &  0.32   & 011 & 0.09  & 0.07  \\
110  &  0.09   &  0.07   & 010 & 0.00  & 0.03 \\
101  &  0.00   &  0.03   & 001 & 0.09  & 0.08  \\
100  &  0.09   &  0.08   & 000 & 0.32  & 0.32  \\
\hline
\hline
\end{tabular}
\end{center}
\caption{The frequencies of length-3 sequences obtained from the Example
  B and the \eM\ reconstructed at $\Range = 3$.
  } 
\label{tab:word_probs_3B}
\end{table}

\begin{table}
\begin{center}
\begin{tabular}{lcc||lcc}
\hline \hline
Sequence   & Example B & \eMSR\ & Sequence   & Example B & \eMSR\ \\
\hline
1111  &  0.23   &  0.29   & 0111 & 0.09  & 0.03  \\
1110  &  0.09   &  0.03   & 0110 & 0.00  & 0.04  \\
1101  &  0.00   &  0.03   & 0101 & 0.00  & 0.00  \\
1100  &  0.09   &  0.04   & 0100 & 0.00  & 0.03  \\
1011  &  0.00   &  0.03   & 0011 & 0.09  & 0.05  \\
1010  &  0.00   &  0.00   & 0010 & 0.00  & 0.03  \\
1001  &  0.00   &  0.05   & 0001 & 0.09  & 0.03   \\
1000  &  0.09   &  0.03   & 0000 & 0.23  & 0.29   \\
\hline
\hline
\end{tabular}
\end{center}
\caption{The frequencies of length-4 sequences obtained from the Example
  B and the \eM\ reconstructed at $\Range = 3$.
  } 
\label{tab:word_probs_4B}
\end{table}

We find by direct calculation that the Example B process has a configurational
entropy of $\hmu = 0.51$ bits/spin, a statistical complexity of $\Cmu = 2.86$
bits, and an excess entropy of $\EE = 0.82$ bits. The reconstructed process
gives similar results with a configurational entropy $\hmu = 0.54$ bits/spin,
a statistical complexity of $\Cmu = 2.44$ bits, and an excess entropy of
$\EE = 0.83$ bits.

\subsection{Example C}

\begin{figure}
\begin{center}
\resizebox{!}{1.00in}{\includegraphics{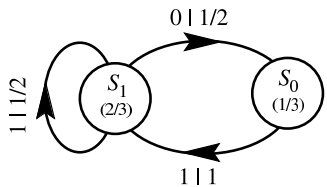}}
\end{center}
\caption{
  The recurrent portion of the \eM\ for the golden mean process, Example C. 
  The process has a memory length of $\Range = 1$, and so
  we label each CS by the last spin seen. 
  }
\label{fig:gm.process}
\end{figure}

We treat this next system, Example C, to contrast it with the last and to
demonstrate how pasts with equivalent futures are merged to form CSs. The
\eM\ for this system is shown in Fig.~\ref{fig:gm.process} and is known as
the {\em golden mean process}. The rule for generating the golden mean process
is simply stated: a 0 or 1 are allowed with equal probability unless the
previous spin was a 0, in which case the next spin is a $1$. Clearly then,
this process needs to only remember the previous spin, and hence it has a
memory length of $\Range = 1$. It forbids the sequence 00 and all sequences
that contain this as a subsequence. The process is so-named because the total
number of allowed sequences grows with sequence length at a rate given by the
golden mean $\phi = (1+\sqrt{5})/2$.  

\begin{figure}
\begin{center}
\resizebox{!}{5cm}{\includegraphics{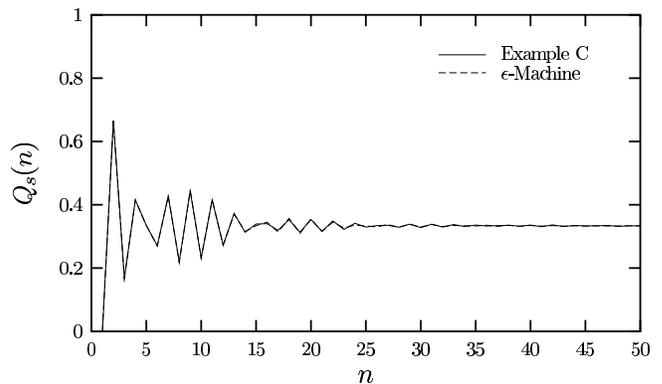}}
\end{center}
\caption{A comparison of the CFs $Q_s(n)$ generated by the $\Range = 1$
  reconstructed \eM\ and the golden mean process of Example C. The CFs
  decay quickly to their asymptotic value of $1/3$.
  }
\label{fig:qs.gm}
\end{figure}

\begin{figure}
\begin{center}
\resizebox{!}{5cm}{\includegraphics{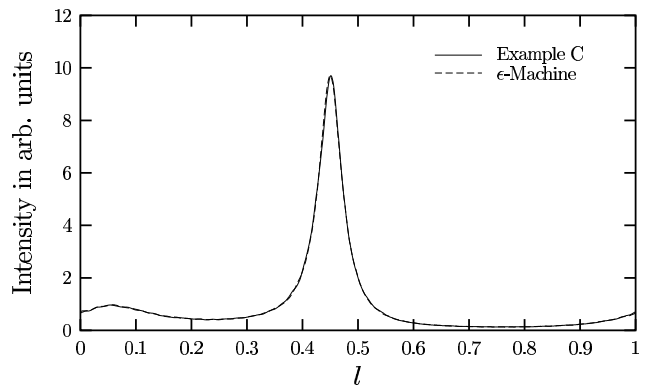}}
\end{center}
\caption{A comparison of the diffraction spectra for Example C and the
  reconstructed $\Range = 1$ \eM. The agreement is excellent. One finds
  a profile ${\cal R}$-factor of 2\% between the ``experimental'' spectrum,
  Example C, and the ``theoretical'' spectrum calculated from the reconstructed
  \eM. 
  } 
\label{fig:diff.gm}
\end{figure}

We employ the \eMSR\ algorithm and find the \eM\ given (again) in
Fig.~\ref{fig:gm.process} at $\Range = 1.$ A comparison of the CFs from
Example C and the golden mean process are given in Fig.~\ref{fig:qs.gm}.
The differences are too small to be seen. We next compare the diffraction
spectra, and these are shown in Fig.~\ref{fig:diff.gm}. We find excellent
agreement and calculate a profile ${\cal R}$-factor of ${\cal R} = 2\%$.
At this point \eMSR\ should terminate, as we have found satisfactory agreement
(to within the numerical error of our technique) between ``experiment'',
Example C, and ``theory'', the reconstructed \eM.  

\begin{figure}
\begin{center}
\resizebox{!}{1.75in}{\includegraphics{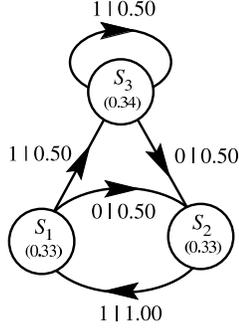}}
\end{center}
\caption{
  The $\Range = 2$ reconstructed nonminimal \eM\ for the golden mean process,
  Example C.
  }
\label{fig:dB.2.gm.process}
\end{figure}

Let us suppose that instead, we increment $\Range$ and follow the
\eMSR\ algorithm as if the agreement at $\Range =1$ had been unsatisfactory.
In this case, we would have generated the ``\eM'' shown in
Fig.~\ref{fig:dB.2.gm.process} at the end of step 3b. We have yet to apply the
equivalence relation Eq. (\ref{eq:cm}) and so let us call this the
\emph{nonminimal} \eM. That is, we have not yet combined pasts with equivalent
futures to form CSs, step 3c. Let us do that now. 

We observe that the state $\CausalState_2$ is different from the other two,
$\CausalState_1$ and $\CausalState_3$, in that one can only see the spin $1$
upon leaving this state. Therefore it cannot possibly share the same futures
as $\CausalState_1$ and $\CausalState_3$, so no equivalence between them is
possible. However, we do see that
$\Prob(1|\CausalState_1) = \Prob(1|\CausalState_3) = 1/2$ and
$\Prob(0|\CausalState_1) = \Prob(0|\CausalState_3) = 1/2$ and, thus, these
states share the same probability of seeing futures of length-$1$. More
formally, we can write 
\begin{eqnarray}
   {\sf T}_{01 \rightarrow 1s}^{(s)} = {\sf T}_{11 \rightarrow 1s}^{(s)} ~.
\label{eq:collapse_nodes}
\end{eqnarray}
Since we are labeling the states by the last two symbols seen at $\Range = 2$,
within our approximation they do have the same futures and thus
$\CausalState_1$ and $\CausalState_3$ can be merged to form a single CS. 
The result is the \eM\ shown in Fig. \ref{fig:gm.process}.

In general, in order to merge two histories, we check that each has an 
equivalent future up to the memory length $\Range$. In this example, we need
only check futures up to length-$1$, because after the addition of one spin
($s$) each is labeled by the same past, namely $1s$. Had we tried to merge
the pasts $11$ and $10$, we would need to check all possible futures after
the addition of {\em two} spins, after which the states would have the same
futures (by assumption). That is, we would require
\begin{eqnarray}
   {\sf T}_{11 \rightarrow 1s}^{(s)} = {\sf T}_{10 \rightarrow 0s}^{(s)} 
\label{eq:collapse_nodes1}
\end{eqnarray}  
and
\begin{eqnarray}
   {\sf T}_{1s \rightarrow ss^{\prime}}^{(s^{\prime})} = {\sf T}_{0s \rightarrow ss^{\prime}}^{(s^{\prime})} 
\label{eq:collapse_nodes2}
\end{eqnarray} 
for all $s,s^{\prime}$.  

We find by direct calculation from the \eM\ that the both Example C and the
reconstructed process have a configurational entropy of $\hmu \approx 0.67$
bits/spin, a statistical complexity of $\Cmu \approx 0.92$ bits, and an
excess entropy of $\EE \approx 0.25$ bits.

\subsection{Example D}

\begin{figure}
\begin{center}
\resizebox{!}{1.00in}{\includegraphics{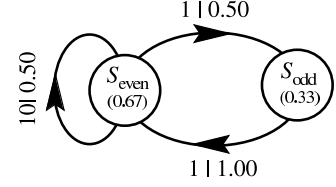}}
\end{center}
\caption{ 
  The recurrent portion of the \eM\ for the even process, Example D.
  Since the CSs cannot be specified by a finite history of previous
  spins, we have labeled them $\CausalState_{\rm even} $ and
  $\CausalState_{\rm odd}$.  
  }
\label{fig:even.process}
\end{figure}

\begin{figure}
\begin{center}
\resizebox{!}{1.75in}{\includegraphics{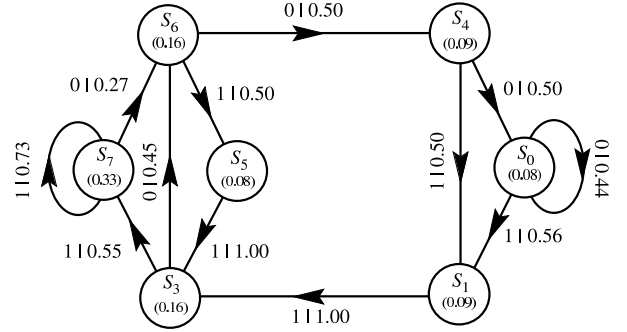}}
\end{center}
\caption{The $\Range = 3$ reconstructed \eM\ for the even process of Example D.
  Since the even process forbids the sequences
  $\{ 01^{2k+1}0, k = 0,1,2,\ldots \}$ and all sequences containing them,
  it is satisfying to see that $010$ is forbidden by the reconstructed \eM,
  as evidenced by the missing $\CausalState_2$ CS.
  }
\label{fig:dB3.example.D}
\end{figure}

We now consider a simple finite-state process that cannot be represented by a
finite-order Markov process, called the \emph{even process},~\cite{cf01,c92}
as the previous examples could. The \emph{even language}~\cite{hu79,bp97}
consists of sequences such that between any two $0$'s either there are no
$1$s or an even number. In a sequence, therefore, if the immediately preceding
spin was a $1$, then the admissibility of the next spin requires remembering
the \emph{evenness} of the number of previous consecutive $1$s, since seeing
the last $0$. In the most general instance, this requires an indefinitely
long memory and so the even process cannot be represented by any finite-order
Markov chain.

We define the even process as follows: If a $0$ or an even number of
consecutive $1$s were the last spin(s) seen, then the next spin is either
$1$ or $0$ with equal probability; otherwise the next spin is $1$. While
this might seem somewhat artificial for the stacking of simple polytypes,
one cannot exclude this class of (so-called \emph{sofic}) structures on
physical grounds. Indeed, such long-range memories may be induced in a
solid-state phase transformations between two crystal
structures.~\cite{kp88,vcup} It is instructive, therefore, to explore the
results of our procedure on processes with such structures.

Additionally, analyzing a sofic process provides a valuable test of
\eMSR\ as practiced here. Specifically, we invoke a finite-order Markov
approximation for the solution of the $\Range = 3$ equations, and we shall
determine how closely this approximates the even process with its
effectively infinite range.  

The \eM\ for this process is shown in Fig. \ref{fig:even.process}. Its
causal-state transition structure is equivalent to that in the \eM\ for the
golden mean process. They differ only in the \emph{spins} emitted upon
transitions out of the $\CausalState_1$ CS. It seems, then, that this
process should be easy to detect.

\begin{figure}
\begin{center}
\resizebox{!}{5cm}{\includegraphics{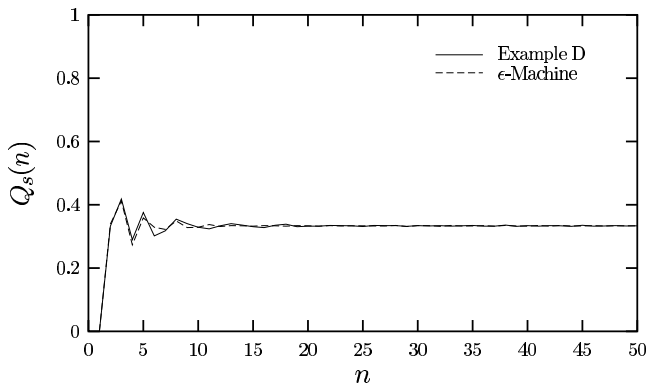}}
\end{center}
\caption{A comparison of the CFs $Q_s(n)$ generated by the $\Range = 3$
  reconstructed \eM\ and the even process of Example D. The CFs decay quickly
  to their asymptotic value of $1/3$.
  }
\label{fig:4}
\end{figure}

The result of \eM\ reconstruction at $\Range = 3$ is shown in
Fig.~\ref{fig:dB3.example.D}. Again, it is interesting to see if the sequences
forbidden by the even process are also forbidden by the $\Range = 3$ \eM. One
finds that the sequence $010$---forbidden by the process---is also forbidden
by the reconstructed \eM. This occurs because CS $\CausalState_2$ is missing. 
We do notice that the reconstructed \eM\ has much more ``structure'' than the
original process. We now examine the source of this additional structure.

Let us first contrast differences between \eMSR\ and other \eM\ reconstruction
techniques, taking the subtree-merging method (SMM) of Crutchfield and
Young~\cite{cy89,c94} as the alternative prototype. There are two major
differences. First, since here we estimate sequence probabilities from the
diffraction spectra and not a symbol sequence, we find it necessary to invoke
the memory-length reduction approximation at $\Range \ge 3$ to obtain a
complete set of equations. Specifically, we assume that only histories up
to range $\Range$ are needed to make an optimal prediction of the next spin.  
Second, we assume that we can label CSs by their length-$\Range$ history. 

We can test these assumptions in the following way. For the first, we 
compare the frequencies of length-$4$ sequences obtained from each method.
This is shown in Table \ref{tab:word_probs}. The agreement is excellent.
All sequence frequencies are within $\pm 0.01$ of the correct values. The
small differences are due to the memory-length reduction approximation.
So this does have an effect, but it is small here.

\begin{figure}
\begin{center}
\resizebox{!}{2.50in}{\includegraphics{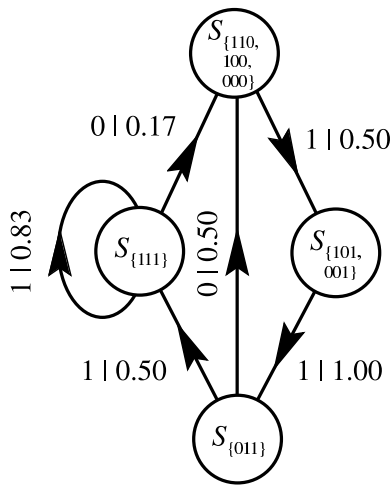}}
\end{center}
\caption{
  The \eM\ inferred from the exact sequence frequencies. The causal states
  are labeled with the (possibly several) histories that can lead to them.
  }
\label{fig:even.exact.process}
\end{figure}

To test the second assumption, we can compare the \eMs\ generated from each
method given the same length-$4$ sequence probabilities. Doing so, SMM gives
the \eM\ for the even process shown in Fig. \ref{fig:even.process}.
\eMSR\ gives a different result. After merging pasts with equivalent futures,
one finds that shown in Fig. \ref{fig:even.exact.process}. For clarity, we
explicitly show the length-$3$ sequence histories associated with each CS,
but do not write out the asymptotic state probabilities.

The \eM\ generated by \eMSR\ is in some respects as good as that generated by
SMM. Both reproduce the sequence probabilities up to length-$4$ from which
they were estimated. The difference is that for \eMSR, our insistence that
histories be labeled by the last $\Range$-spins forces the representation to be 
Markovian of range $\Range$. Here, a simpler model for the process, as
measured by the smaller statistical complexity ($0.92$ bits as compared to
$1.92$ bits), can be found. So the notion of minimality is violated. That is,
\eMSR\ searches only a subset of the space from which processes can belong.
Should the true process lie outside this subset (Markovian processes of range
$\Range$), then \eMSR\ returns an approximation to the true process. The
approximation may be both more more complex and less predictive than the true
process. It is interesting to note that had we given SMM the sequence
probabilities found from the solutions of the spectral equations, we would
have found, (within some error) the \eM\ given in Fig. \ref{fig:even.process}.

We find, then, that there are two separate consequences to applying
\eMSR\ that affect the reconstructed \eM. The first is that for $\Range \ge 3$,
the memory-length reduction approximation must be invoked to obtain a complete
set of equations. This approximation limits the histories treated and can
affect the values estimated for the sequence probabilities. The second is the
state-labeling scheme. Only for Markovian (non-sofic) processes can CSs be
labeled by a unique finite history. Making this assumption effectively limits
the class of processes one can detect to those that are block-$\Range$
Markovian. To see this more clearly, we can catalog the possible histories
that lead to the two CSs in Fig. \ref{fig:even.process}. In doing so, we find
that the histories $000$, $011$, $110$, $100$, and $100$ always leave the 
process in CS $\CausalState_{\rm even}$. Similarly, the histories $001$ and
$101$ always leave the process in CS $\CausalState_{\rm odd}$. But having
seen the history $111$ does not specify the CS as one can arrive in both CSs
from this history. So the labeling of CSs by histories fails here. 

Then why do we not find sequence probabilities by solving the spectral equations
and then use SMM to reconstruct the \eM? There are two reasons. The first is
that in general one must know sequence probabilities for longer sequences than
is necessary for \eMSR. Solving the spectral equations for these longer
sequence frequencies is onerous. The second is that error in the sequence
probabilities found from solving the spectral equations for these longer
sequences makes identifying equivalent pasts almost impossible. The even
process is an exception here, since one needs to consider only futures of
length $1$. This is certainly not the case in general.

\begin{table}
\begin{center}
\begin{tabular}{lcc||lcc}
\hline \hline
Sequence   & \eMSR\ & SMM & Sequence   & \eMSR\ & SMM \\
\hline
1111  &  0.24   &  0.25   & 0111 & 0.08  & 0.09  \\  
1110  &  0.09   &  0.08   & 0110 & 0.07  & 0.08  \\   
1101  &  0.09   &  0.08   & 0101 & 0.00  & 0.00  \\   
1100  &  0.08   &  0.08   & 0100 & $<$ 0.01  & 0.00  \\ 
1011  &  0.08   &  0.08   & 0011 & 0.08  & 0.08  \\  
1010  &  0.00   &  0.00   & 0010 & $<$ 0.01 & 0.00  \\  
1001  &  0.04   &  0.04   & 0001 & 0.05  & 0.04   \\  
1000  &  0.04   &  0.04   & 0000 & 0.04  & 0.04   \\ 
\hline
\hline
\end{tabular}
\end{center}
\caption{The frequencies of length-$4$ sequences obtained from \eMSR\ and
  SMM for the even process. At most, they differ by $\pm 0.01$.
  }
\label{tab:word_probs}
\end{table}

\begin{figure}
\begin{center}
\resizebox{!}{5cm}{\includegraphics{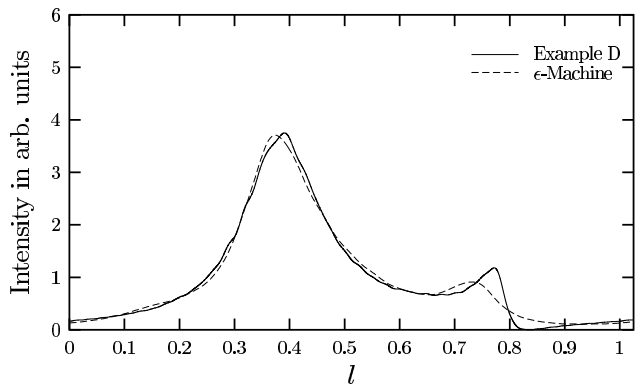}}
\end{center}
\caption{A comparison between the diffraction spectra $\mathsf{I}(l)$
  generated by the $\Range = 3$ reconstructed \eM\ and by the even process of
  Example D. The agreement is good except in the region $ 0.7 < l < 0.9$.
  Notably, the diffraction spectra for the even process has an isolated zero
  at $l = 5/6$.
  }
\label{fig:dp.example.D}
\end{figure}

Having explored the differences between \eMSR\ and SSM, we now return to a
comparison of the results of each method. A comparison of the CFs for the even
process and the reconstructed \eM\ is given in Fig.~\ref{fig:4}. We see that
both decay quite quickly to their asymptotic values of $1/3$. There is good
agreement, except in the region between $5 \leq n \leq 10$. Examining the
diffraction spectra in Fig.~\ref{fig:dp.example.D}, we see that there is
likewise good agreement except in the region $0.7 < l < 0.9$. The profile
${\cal R}$-factor between the two spectra is ${\cal R} \approx 8\%$, which
indicates that there is reasonable agreement. 

There is a curious isolated zero in the process's spectrum at $l = 5/6$. The
other interesting feature is the broad peak at $l \approx 1/3$. One might
guess that this originates from some 3C$^+$ structure and, indeed, glancing
at the reconstructed \eM\ of Fig.~\ref{fig:dB3.example.D} shows that the
CSC [$\CausalState_7$] is strongly represented. The faulting is less clear.
We would expect, though, that presence of the CSC
[$\CausalState_7 \CausalState_6 \CausalState_4 \CausalState_0 \CausalState_1 \CausalState_3$]
would indicate layer-displacement faulting and the CSC
[$\CausalState_7 \CausalState_6 \CausalState_5 \CausalState_3$] supports
this.

We find by direct calculation from the even process that it has a
configurational entropy of $\hmu = 0.67$ bits/spin, a statistical complexity
of $\Cmu = 0.92$ bits, and an excess entropy of $\EE = 0.91$ bits. The
reconstructed \eM\ gives information-theoretic quantities that are rather
different. We find a configurational entropy $\hmu = 0.80$ bits, a statistical
complexity of $\Cmu = 2.63$ bits, and an excess entropy of $\EE = 0.22$ bits.

One reason that the reconstructed \eM\ gives CFs and diffraction spectra in
such good agreement with the even process in spite of the fact that the
information-theoretic quantities are different is the insensitivity of the
CFs and diffraction spectra to the probabilities of long sequences:
Eq.~(\ref{eq:Q.to.p}) adds sequence probabilities to find CFs. The fact that
the even process has such a long memory is masked by this. However,
information-theoretic quantities are sensitive to the structure of long
sequences. \EM\ reconstruction at $\Range = 4$ should prove interesting,
in this light, since the even process picks up another forbidden
sequence---$01110$---and this additional structure would be reflected in
the reconstructed \eM.

For comparison we list each of the example's information-theoretic
properties in Table \ref{tab:results}.

\begin{table}
\begin{center}
\begin{tabular}{lccccc}
\hline \hline
System   & Range & $\hmu$ [bits/ML] & $\Cmu$ [bits] & $\EE$ [bits] &$\Delta$\\
\hline
Example A& 3     &  0.44              &  2.27         &  0.95    & 0.00    \\
\eM\     & 3     &  0.44              &  2.27         &  0.95    & 0.00    \\
\hline
Example B& 4     &  0.51              &  2.86         &  0.82    & 0.00    \\
\eM\     & 3     &  0.54              &  2.44         &  0.83    &-0.01    \\
\hline
Example C& 1     &  0.67              &  0.92         &  0.25    & 0.00    \\
\eM\     & 1     &  0.67              &  0.92         &  0.25    & 0.00    \\
\hline
Example D&$\infty$& 0.67              &  0.92         &  0.91    &         \\
\eM\     & 3     &  0.79              &  2.58         &  0.21    & 0.00    \\
\hline
\hline
\end{tabular}
\end{center}
\caption{Measures of intrinsic computation calculated from the processes of
  Examples A, B, C, and D and their ($\Range = 3$) reconstructed \eMs. For
  Examples A, B, and C the reconstructed \eMs\ give good agreement. For Example
  D, however, the reconstructed \eM\ requires more memory and still has
  a entropy density $\hmu$ significantly higher than that of the even process.
  The last column gives $\Delta = \Cmu - \EE - \Range \hmu$ as a consistency
  check derived from Eq.~(\ref{BlockMarkovComplexityReln}), which describes
  order-$\Range$ Markov processes. Recall that the even process of Example D 
  is not a finite-$\Range$ process and so Eq.~(\ref{BlockMarkovComplexityReln})
  does not hold. All one can say is that $\EE \leq \Cmu$, which is
  the case for Example D.
  }
\label{tab:results}
\end{table}

\subsection{Possible Difficulties with Applying \eMSR}

We have given four examples that demonstrate successful applications of \eMSR.  
We have found instances, however, when the \eMSR\ has difficulties converging
to a satisfactory result. We now analyze each step in \eMSR\ as given in
Table~\ref{SpectraleMReconstruction} and discuss possible problems that may
be encountered. 

\begin{trivlist}

\item {\it Step 1}. 
Several problems can arise here. One is that the figures-of-merit, $\beta$
and $\gamma$, are sufficiently different from their theoretical values over
all possible $l$-intervals that \eMSR\ should not even be attempted. Even
if one does find an interval such that they indicate satisfactory spectral
data, it is possible that the CFs extracted over this interval are unphysical.
That is, there is no guarantee that all of the CFs are both positive and
less than unity. In such a case, no stacking of MLs can reproduce these CFs. 
Finally, if error ranges have not been reported with the experimental data,
it may not be possible to set the error threshold $\Gamma$. 

\item {\it Step 2}. 
The $\Prob(\omega^\Range)$ solutions to the spectral equations are not
guaranteed to be either real or positive for $\Range \geq 3$. If this is so,
then no physical stacking of MLs can reproduce the CFs from the spectrum. 

\item {\it Step 3}.
Given $\Prob(\omega^\Range)$ that satisfy the elementary conditions of
probability ({\it i.e.}, there is no difficulty at step 2), step 3 will
return a machine that generates $\Prob(\omega^\Range)$. It is possible,
however, that the resulting states are not \emph{strongly connected}, and
thus the result may not be interpreted as a single \eM. 

\item {\it Step 4}.
There are no difficulties here.

\item {\it Step 5}.
It is possible that one is required to go to an $\Range$ that is cumbersome
to calculate. In this case, one terminates the procedure through practicality.

\end{trivlist}

We find that the roots of these difficulties can be ultimately traced to four
problems: (i) excessive error in the diffraction spectrum, (ii) the process
has statistics that are too complex to be captured by a finite-range Markov
process, (iii) the memory-length approximation is not satisfied, and
(iv) the initial assumptions of polytypism are violated. We are likely to
discover (i) in step 1. For (ii) and (iii), we find no difficulties at step 1,
but rather at steps 2, 3, and 5. For (iv), we have not examined this case in
detail. However, we expect that if the assumptions of the stacking of MLs
(\S \ref{EMFaultModelStructuralAnalyses}) are not met, then since Eq.
(\ref{eq:diff1}) is no longer valid, the CFs found by Fourier analysis will
not reflect the actual stacking probabilities. This will likely be interpreted
as poor figures-of-merit, and \eMSR\ will terminate at step 1. 
 
Of the three possible difficulties only (ii) and (iii) should be considered to
be inherent to \eMSR. It is satisfying that \eMSR\ can detect errors in the
diffraction spectrum and then stop, so that it does not generate an invalid
representation that simply describes ``error'' or ``noise''. 

\section{Characteristic Lengths in CPSs} 
\label{CPSCharacteristicLengths} 

We now return to one of the mysteries of polytypism, namely that of the
long-range order which they seem to possess. It is of interest, then, to
ask what, if anything, the spectrally reconstructed \eM\ indicates about the
range of interactions between MLs. In this section, we discuss and quantify
several characteristic lengths that can be estimated from reconstructed \eMs.

\begin{trivlist}

\item ({\it i}) \emph{Correlation Length,} $\lambda_c$. From statistical
mechanics, we have the notion of a correlation length,~\cite{bdfn93,y92}
which is simply the characteristic length scale over which ``structures'' are
found.  The correlation functions $Q_c(n)$, $Q_a(n)$, and $Q_s(n)$
are known to decay to $1/3$ for many disordered stackings.
(For some exceptions, see Kabra and Pandey,~\cite{kp88} Yi
and Canright,~\cite{yc96} and Varn.~\cite{v01}) For the disordered cases
considered here, exponential decay to $1/3$ seems to be the
rule.  We therefore define the \emph{correlation length} $\lambda_c$ as
the characteristic length over which correlation information is lost
with increasing separation $n$. More precisely, let us define $\Psi_q(n)$ as
\begin{eqnarray}
   \Psi_q(n) = \sum_{\alpha} \Bigl|Q_{\alpha}(n) -\frac{1}{3}\Bigr| ~,
\label{eq:def_Psi}
\end{eqnarray}
so that $\Psi_q(n)$ gives a measure of the deviation of the CFs from their
asymptotic value. Then we say that
\begin{eqnarray}
   \Psi_q(n) = F(n) \times 2^{-n/\lambda_c} ~,
\label{eq:def_corr_length}
\end{eqnarray}
where $F(n)$ is some function of $n$.

For those cases where the CFs do not decay to $1/3$, we say that the
correlation length is infinite. We find that exponential decay is not
always obeyed, but it seems to be common,\cite{note2} 
and the correlation length thus defined gives a useful measure of the
rate of coherence loss as $n$ increases. Our definition of correlation
length is similar to the \emph{characteristic length} $L$ defined by
Shrestha and Pandey.~\cite{sp96,sp97}

\item ({\it ii}) \emph{Recurrence Length,} $\mathcal{P}$. For an exactly
periodic process, the period gives the length over which a template pattern
repeats itself. We can generalize this for arbitrary, aperiodic processes
in the following way. Let us take the \emph{recurrence length}
$\mathcal{P}$ as the geometric mean of the distances between visits to
each CS weighted by the probability to visit that CS:
\begin{eqnarray}
   \mathcal{P} \equiv \prod_{\CausalState_i \in \CausalState} T_i^{p_i} ~,
\label{eq:def.recurrence.length}
\end{eqnarray}
where $T_i$ is the average distance between visits to a CS
and $p_i$ is the probability of visiting that CS. Then,
\begin{eqnarray}
\mathcal{P} & = & \prod_{\CausalState_i \in \CausalState} (2^{\log_2
   T_i})^{p_i}  \nonumber \\
  & = & \prod_{\CausalState_i \in \CausalState} 2^{- p_i \log_2 p_i}  \nonumber \\
  & = & 2^{- \sum_{\CausalState_i \in \CausalState} p_i \log_2 p_i}  \nonumber \\
  & = & 2^{\Cmu} ~,
\label{eq:def.recurrence.length1}
\end{eqnarray}
where we have used the relation $T_i = 1/p_i$.

For periodic processes, $\Cmu = \log_2\mathcal{P}$ and so $\mathcal{P}$ is
simply a process's period. For aperiodic processes $\mathcal{P}$ gives a
measure of the average distance over which the \eM\ returns to a CS. Notice
that this is defined as the average recurrence length \emph{in the H{\"{a}}gg
notation}. For cubic and rhombohedral structures, for example, this is
one-third of the physical repeat distance in the absolute stacking sequence.

\item ({\it iii}) \emph{Memory Length,} $\MLength$. Recall from \S
\ref{MeasuresStructureIntrinsicComputation} that the \emph{memory length} is
an integer which specifies the maximum number of previous spins that one
must know in the worst case to make an optimal prediction of the next spin.
For an $\Range^{\mathrm {th}}$-order Markov process this is $\Range$.

\item ({\it iv}) \emph{Interaction Length,} $\InteractionLength$. The
\emph{interaction length}
is an integer that gives the maximum range over which spin-spin interactions
appear in the Hamiltonian.

\end{trivlist}

We calculated the $\lambda_c$, $\mathcal{P}$, and $\MLength$ (in units of MLs)
for Examples A-D as well as for three crystal structures. The results are
displayed in Table \ref{tab:lengths}. We see that each captures a different
aspect of the system. The correlation length $\lambda_c$ sets a scale over
which a process is coherent. For crystals, as shown in Table \ref{tab:lengths},
this length is infinite. For more disordered systems, this value decreases. 
The generalized period $\mathcal{P}$ is a measure of the scale over which the
pattern produced by the process repeats. The memory length $\MLength$ is most
closely related to what we might think as the maximum range of ``influence''
of a spin. That is, it is the maximum distance over which one might need to
look to obtain information to predict a spin's value.

For periodic, infinitely correlated systems spins at large separation carry
information about each other, as seen in crystals. But this information is
redundant. Outside a small neighborhood one gets no additional information
by knowing the orientation a spin assumes. Notice that one can have an infinite 
memory length with a relatively small correlation length, as seen for the even
system (Example D). That is, even though on \emph{average} the knowledge one
has about a spin may decay, there are still configurations in which distantly 
separated spins carry information about each other that is not stored in the
intervening spins.  

If we know the \eM\ for a process, then we can directly calculate $\lambda_c$,
$\mathcal{P}$, and $\MLength$. How, then, do these relate to the interaction
length $\InteractionLength$? Infinite correlation lengths can be achieved with
very small $\InteractionLength$, as in the case of simple crystals.
So correlation lengths alone imply little about the range of interactions. 
For a periodic system in the ground state, the configuration's period
puts a lower bound on the interaction length via
$\InteractionLength \geq \log_2 \mathcal{P}$---barring fine tuning
of parameters, such as found at the multiphase boundaries in the
ANNNI model~\cite{y88} or those imposed by symmetry
considerations.~\cite{vc01,cw96} This does not hold, however, for systems
above the ground state. The most likely candidate for a useful relation
between $\InteractionLength$ and a quantity generated from the \eM\ is
$\MLength$. Again, $\MLength$ sets a lower bound on $\InteractionLength$,
\emph{if} the system is in the ground state. For polytypes, the multitude of
observed structures suggests that most are not in the ground state and, thus,
one does not know what the relation between $\InteractionLength$ and $\MLength$
is. It is conceivable, especially in the midst of a solid-state phase
transition, that small $\InteractionLength$ could generate large $\MLength$.
Although, an \eM\ is a complete description of the underlying stacking process,
one should be able to calculate from it the interaction length in the case
that the system not in the ground state. We suspect the answer lies in the
different ways in which a Hamiltonian and an \eM\ describe a material.

\begin{table}
\begin{center}
\begin{tabular}{lccc}
\hline
\hline
 System     \hspace{6mm}&  \hspace{4mm} $\lambda_c$ \hspace{4mm}  &
 \hspace{4mm} $\mathcal{P}$ \hspace{4mm} & \hspace{4mm} $\MLength$ 
 \hspace{4mm} \\
\hline
Example A, $\Range =3$     &  $\sim 7.4$   &  4.8  & 3        \\
Example B, $\Range =4$     &  $\sim 7.8$   &  7.3  & 4        \\
Example C, Golden Mean     &  $\sim 4.5$   &  1.9  & $1$ \\
Example D, Even Process    &  $\sim 1.7$   &  1.9  & $\infty$ \\
3C            &  $\infty$     &  1    & 0        \\
2H            &  $\infty$     &  2    & 1        \\
6H            & $\infty$      &  6    & 3        \\
\hline
\hline
\end{tabular}
\end{center}
\caption{A comparison of the three characteristic lengths that one can
  calculate from knowledge of the \eM: the correlation length $\lambda_c$,
  the recurrence length $\cal P$, and the memory length $\MLength$.
  }
\label{tab:lengths}
\end{table}

\section{Conclusions}
\label{Conclusions}

We offer here a new way of treating planar disorder in CPSs. We find the
minimal, unique description of the stacking for any amount and kind of
ordered and disordered sequence that a material may contain. We demonstrated
how this description---the \eM---can be directly inferred from experimental
diffraction spectra. \eMSR\ uses all of the information in the spectrum, both
Bragg-like and diffuse scattering.  Our description is necessarily
statistical, in that we do not find the specific stacking sequence that
gave rise to the diffraction pattern, but rather a minimal description
of the ensemble of stackings that could have generated the diffraction
spectrum.  We contend that this statistical description is the most useful
form in which to express the structure of the crystal. From it, physical
parameters such as the entropy per ML, the statistical complexity, and the
average stacking-fault energy for disordered stacking sequences are
calculable.~\cite{vcc02c} Indeed, could we have found a specific stacking
sequence millions of MLs in length, one still would search for some way to
compress this information into a useful form. In short, one would want to
find its \eM. We therefore state that our solution---which is directly
determined from experimental diffraction spectra---offers the most complete and
compact understanding of the stacking process and, thus, of the solid's
structure. On this basis, we say that ({\it i}) \eMSR\ has solved the problem
of inferring structural information from diffraction spectra for CPSs and
({\it ii}) the resulting \eM\ is the unique and minimal expression of that
structure.

We illustrated the approach by solving the inference problem for $\Range = 3$,
which gives stacking-sequence probabilities up to length $4$. Unlike previous
work, we considered \emph{all} the possible sequences of this length without
invoking symmetry arguments. This highlights the need for good experimental
data. It is important to get accurate diffraction spectra over a unit
$l$-interval that records the intensity of both the Bragg-like peaks and the
diffuse scattering. We have shown that there exist quantities, called the
figures-of-merit, that measure the quality of the spectral data over a
particular unit $l$-interval and that help determine a suitable interval for
\eM\ spectral reconstruction.

While we have addressed only CPSs here, the extension to other layered
structures is straightforward. A first theoretical task in this is to
find an expression for the diffracted intensity in terms of suitable CFs
and to relate these CFs to the sequence distribution (and thence to an
\eM). While such an \eM\ may draw from an alphabet larger than two for more
complicated polytypes, such as micas and kaolins,~\cite{vc01} there are in
principle no theoretical obstacles to applying \eMSR\ to more complicated
polytypic structures. 

More generally, \eMSR\ also contributes to the machine-learning side of
computational mechanics. \eMSR\ is novel, in that we use a power spectrum
to reconstruct the \eM\ instead of a temporal data sequence, as prior
algorithms have. We see this as a prelude to the question of how one infers
an \eM\ from general spectral data and are continuing research along
these lines.

There are, however, some limitations to \eMSR, as presented here. We only
attempted \eM\ reconstruction up to $\Range = 3$. While in principle one can
attempt it for any $\Range$, there are computational complexity difficulties.
In the most general case, the number of variables one needs to solve
for is exponential in $\Range$, and many of the equations are nonlinear.
More seriously, the maximum number of terms in any equation grows as an
exponential of an exponential in $\Range$.  For $\Range = 3$, there were $11$
terms in two of the equations. At $\Range = 4$, two of the equations have
$171$ terms, all of them nonlinear. For $\Range = 5$, this grows to $43 690$
terms.~\cite{v01} These terms are all additive, so a fortuitous cancellation
is not possible. It is possible, however, that physical insight into the
relative importance of sequences may allow one to neglect a number of
terms in these equations. We feel that the general case of $\Range = 4$
is tractable, and this is a subject of current research. We also
suspect that there are alternative algorithms that will greatly reduce
the computational complexity of finding solutions.

Finally, we stress that there is a difference between structure and mechanism
in disordered stacking sequences. The \eM\ describes the structure, but has
little to say about how the material came to be stacked in this fashion.
While it is possible to formally identify CSCs with ``faulting structures'',
this can be misleading. It is certainly possible that the cumulative effects
of repeated faulting by a particular mechanism may lead to a structure that
is different from a crystal simply permeated with that kind of fault. That
is, for high fault densities, adjacent faults may be produced in the same way,
but the close proximity of the faults may cause us to interpret the structure
differently---e.g., as a small segment of complex crystal.  

In order to determine the mechanism of faulting in, say, an annealed crystal
undergoing a solid-state phase transition, it is desirable to begin with many
(identical) crystals and arrest the solid-state transformation at various
stages. By reconstructing the \eM\ after different annealing times, the
route to disorder can be made plain. The result is a picture of how structure
(as captured by intermediate \eMs) changes during annealing. This change in
structure should give direct insight into the structure-forming mechanisms.
This should be compared with the numerical simulation of faulting in a crystal,
such as those done by Kabra and Pandey,~\cite{kp88} Engel,~\cite{e90} Shrestha
and Pandey,~\cite{sp96,sp97} Gosk,~\cite{g00,g01} and Ramasesha and
Rao.~\cite{rr77} We note that in such simulations, the \eM\ can be directly
calculated from the sequence to high accuracy. Some experimental work on
solid-state phase transitions has been done,~\cite{sk94} but we hope that
there will be additional effort in this direction.

\begin{acknowledgments}

This work was supported at the Santa Fe Institute under the Networks
Dynamics Program funded by the Intel Corporation and under the Computation,
Dynamics and Inference Program via SFI's core grants from the National
Science and MacArthur Foundations. Direct support was provided by NSF
grants DMR-9820816 and PHY-9910217 and DARPA Agreement F30602-00-2-0583.
DPV's visit to SFI was partially supported by the NSF. 

\end{acknowledgments}

\appendix

\section{\label{secAppendix:level1}The Spectral Equations}
\label{SpectralEquations}

\subsection{$\Range = 1$}

The spectral equations at $\Range = 1$ are linear and admit an analytical
solution. Specifically, we write out Eqs. (\ref{eq:cons.of prob.1}),
(\ref{eq:cons.of prob.2}), and (\ref{eq:Q.to.p}) for $\Range = 1$ and
solve them. We find, 
\begin{eqnarray*}
   \Prob(11)& = &Q_r(2) ~,    \\
   \Prob(01)& = &p(10)   =  \frac{1}{2}[1-Q_c(2)-Q_r(2)] ~, \\
   \Prob(00)& = &Q_c(2) ~.
\label{eq:solutions_r_1}
\end{eqnarray*}

\subsection{$\Range = 2$}

Similarly, the spectral equations at $\Range = 2$ are linear and also can be
solved analytically. Again, we write out Eqs. (\ref{eq:cons.of prob.1}),
(\ref{eq:cons.of prob.2}), and (\ref{eq:Q.to.p}) for $\Range = 2$ and
solve them. We find,  
\begin{eqnarray*}
\Prob(000) & = & [ 3Q_c(2) - 2Q_c(3) - 3Q_r(2) - 4Q_r(3) + 3]/6 ~, \\
\Prob(001) & = & [ 3Q_c(2) + 2Q_c(3) + 3Q_r(2) + 4Q_r(3) - 3]/6 ~, \\
\Prob(010) & = & [-3Q_c(2) - 2Q_c(3) - 3Q_r(2) -  Q_r(3) + 3]/3 ~, \\
\Prob(011) & = & [ 3Q_c(2) + 4Q_c(3) + 3Q_r(2) + 2Q_r(3) - 3]/6 ~, \\
\Prob(100) & = & [ 3Q_c(2) + 2Q_c(3) + 3Q_r(2) + 4Q_r(3) - 3]/6 ~, \\
\Prob(101) & = & [-3Q_c(2) -  Q_c(3) - 3Q_r(2) - 2Q_r(3) + 3]/3 ~, \\
\Prob(110) & = & [ 3Q_c(2) + 4Q_c(3) + 3Q_r(2) + 2Q_r(3) - 3]/6 ~, \\
\Prob(111) & = & [-3Q_c(2) - 4Q_c(3) + 3Q_r(2) - 2Q_r(3) + 3]/6 ~.  
\label{eq:solutions_r_2}
\end{eqnarray*}

\subsection{$\Range = 3$}

At $\Range = 3$, we require $16$ relations to constrain the length-$4$
binary-sequence probabilities. Now, however, we encounter nonlinearities,
and by necessity the spectral equations are solved numerically. 
We write them out here. 

At $\Range = 3$, Eq. (\ref{eq:cons.of prob.1}) implies the following seven equations.  

\begin{eqnarray*}
        \Prob(0111) & = & \Prob(1110) ~,                    \\
        \Prob(0001) & = & \Prob(1000) ~,                   \\
   \Prob(0011)  +  \Prob(1011) & = & \Prob(0111)  + \Prob(0110) ~, \\
   \Prob(0101)  +  \Prob(1101) & = & \Prob(1011)  + \Prob(1010) ~, \\
   \Prob(0010)  +  \Prob(1010) & = & \Prob(0101)  + \Prob(0100) ~, \\
   \Prob(0001)  +  \Prob(1001) & = & \Prob(0011)  + \Prob(0010) ~, \\
   \Prob(0100)  +  \Prob(1100) & = & \Prob(1001)  + \Prob(1000) ~. 
\label{eq:eq:cons.of prob.1 r=3}
\end{eqnarray*}
Equation (\ref{eq:cons.of prob.2}) provides for normalization, providing one
additional constraint. Finally, the remaining $8$ relations are obtained by
relating sequence probabilities to CFs as prescribed by Eq. (\ref{eq:Q.to.p}).
We further reduce the last two relations which involve sequence probabilities
of length-$5$ to those of length-$4$ via relations of the form given by
Eq. (\ref{eq:p.reduction}). We find, 
\begin{eqnarray*}
  Q_c(2) &= &\Prob(0000) + \Prob(0001) + \Prob(0010) + \Prob(0011)  ~,  \\ 
  Q_r(2) &= &\Prob(1100) + \Prob(1101) + \Prob(1110) + \Prob(1111)  ~,  \\ 
  Q_c(3) &= &\Prob(0110) + \Prob(0111) + \Prob(1010) + \Prob(1011)      \\ 
         &  & + \Prob(1100) + \Prob(1101)  ~,                   \\ 
  Q_r(3) &= &\Prob(0010) + \Prob(0011) + \Prob(0100) + \Prob(0101)   \\
         &  & + \Prob(1000) + \Prob(1001)  ~,                   \\
  Q_c(4) &= &\Prob(1111) + \Prob(1000) + \Prob(0100) + \Prob(0010)    \\
         &  & + \Prob(0001)  ~,                              \\
  Q_r(4) &= &\Prob(0000) + \Prob(0111) + \Prob(1011) + \Prob(1101)   \\
         &  & + \Prob(1110)  ~,                              \\
  Q_c(5) &= &  \frac{\Prob^2(0000)}{\Prob(0000) + \Prob(0001)}        +\frac{\Prob(0011)\Prob(0111)}{\Prob(0111)+\Prob(0110)}  \\
         &  & +\frac{\Prob(0101)\Prob(1011)}{\Prob(1011)+\Prob(1010)} +\frac{\Prob(0110)\Prob(1101)}{\Prob(1101)+\Prob(1100)}  \\
         &  & +\frac{\Prob(0111)\Prob(1110)}{\Prob(1110)+\Prob(1111)} +\frac{\Prob(1001)\Prob(0011)}{\Prob(0011)+\Prob(0010)}  \\
         &  & +\frac{\Prob(1010)\Prob(0101)}{\Prob(0101)+\Prob(0100)} +\frac{\Prob(1011)\Prob(0110)}{\Prob(0110)+\Prob(0111)}  \\
         &  & +\frac{\Prob(1100)\Prob(1001)}{\Prob(1001)+\Prob(1000)} +\frac{\Prob(1101)\Prob(1010)}{\Prob(1010)+\Prob(1011)}  \\
         &  & +\frac{\Prob(1110)\Prob(1100)}{\Prob(1100)+\Prob(1101)}  ~,                                      \\
  Q_r(5) &= &  \frac{\Prob^2(1111)}{\Prob(1111) + \Prob(1110)}        +\frac{\Prob(1100)\Prob(1000)}{\Prob(1000)+\Prob(1001)}  \\
         &  & +\frac{\Prob(1010)\Prob(0100)}{\Prob(0100)+\Prob(0101)} +\frac{\Prob(1001)\Prob(0010)}{\Prob(0010)+\Prob(0011)}  \\
         &  & +\frac{\Prob(1000)\Prob(0001)}{\Prob(0001)+\Prob(0000)} +\frac{\Prob(0110)\Prob(1100)}{\Prob(1100)+\Prob(1101)}  \\
         &  & +\frac{\Prob(0101)\Prob(1010)}{\Prob(1010)+\Prob(1011)} +\frac{\Prob(0100)\Prob(1001)}{\Prob(1001)+\Prob(1000)}  \\
         &  & +\frac{\Prob(0011)\Prob(0110)}{\Prob(0110)+\Prob(0111)} +\frac{\Prob(0010)\Prob(0101)}{\Prob(0101)+\Prob(0100)}  \\
         &  & +\frac{\Prob(0001)\Prob(0011)}{\Prob(0011)+\Prob(0010)}  ~. 
\label{eq:p.reduction r=3}
\end{eqnarray*}

\bibliography{ipdcps}

\begin{thebibliography}{110}
\bibitem[{not({\natexlab{b}})}]{note3}
\bibinfo{note}{For close-packed crystal structures it is common to use the
  Ramsdell notation, nX, where n gives the total number of ML's in the unit
  cell and X specifies the crystal symmetry. We use H to denote hexagonal
  symmetry, C for cubic, R for rhombohedral and L if the symmetry is unknown.
  This notation is not necessarily unique for some longer period polytypes. We
  will use the Ramsdell notation to specify crystal structures. The equivalents
  in the H{\"{a}}gg notation are given in Table \ref{FMStructuresOneMr_3}. For
  a more complete discussion, see Ref. \onlinecite{sk94}.}
\end{thebibliography}

\end{document}